\documentclass[a4paper,11pt]{article}
\usepackage{jinstpub} % for details on the use of the package, please see the JINST-author-manual
\usepackage{tikz}
\usepackage{graphicx}
\usetikzlibrary{positioning} % Include the positioning library
\usepackage{amsmath}
\usepackage{graphicx} % For including images
\usepackage{caption}
\usepackage{subcaption} % For subfigures
\title{\boldmath Energy Reconstruction of Non-fiducial Electron-Positron Events in the DAMPE Experiment Using Convolutional Neural Networks}

\author[a,1]{Enzo Putti-Garcia,\note{Corresponding author.}}
\author[a]{Andrii Tykhonov,}
\author[a]{Andrii Kotenko,}
\author[a]{Hugo Boutin,}
\author[a]{Manbing Li,}
\author[a]{Paul Coppin,}
\author[a]{Andrea Serpolla,}
\author[b]{Jennifer Maria Frieden,}
\author[b]{Chiara Perrina,}
\author[a]{Xin Wu}
\affiliation[a]{Department of Nuclear and Particle Physics, University of Geneva, rue du Général–Dufour 24, CH-1211 Geneva, Switzerland}
\affiliation[b]{ Institute of Physics, Ecole Polytechnique Fédérale de Lausanne (EPFL), CH-1015, Lausanne, Switzerland}

\emailAdd{enzo.putti-garcia@unige.ch}

\abstract{The Dark Matter Particle Explorer (DAMPE) is a space-based Cosmic-Ray (CR) observatory
with the aim, among others, to study Cosmic-Ray Electrons (CREs) up to 10 TeV. Due to the
low CRE rate at multi-TeV energies, we aim to increasing the acceptance by selecting events outside the fiducial volume. The complex topology of non-fiducial events requires the development of a novel energy reconstruction method. We propose the usage of Convolutional Neural Networks for a regression task to recover an accurate estimation of the initial energy.}


\arxivnumber{2503.10521} % Only if you have one

\begin{document}
\maketitle
\flushbottom

\section{Introduction}
\label{sec:intro}
\begin{figure}[htbp]
\centering
\includegraphics[width=.7\textwidth]{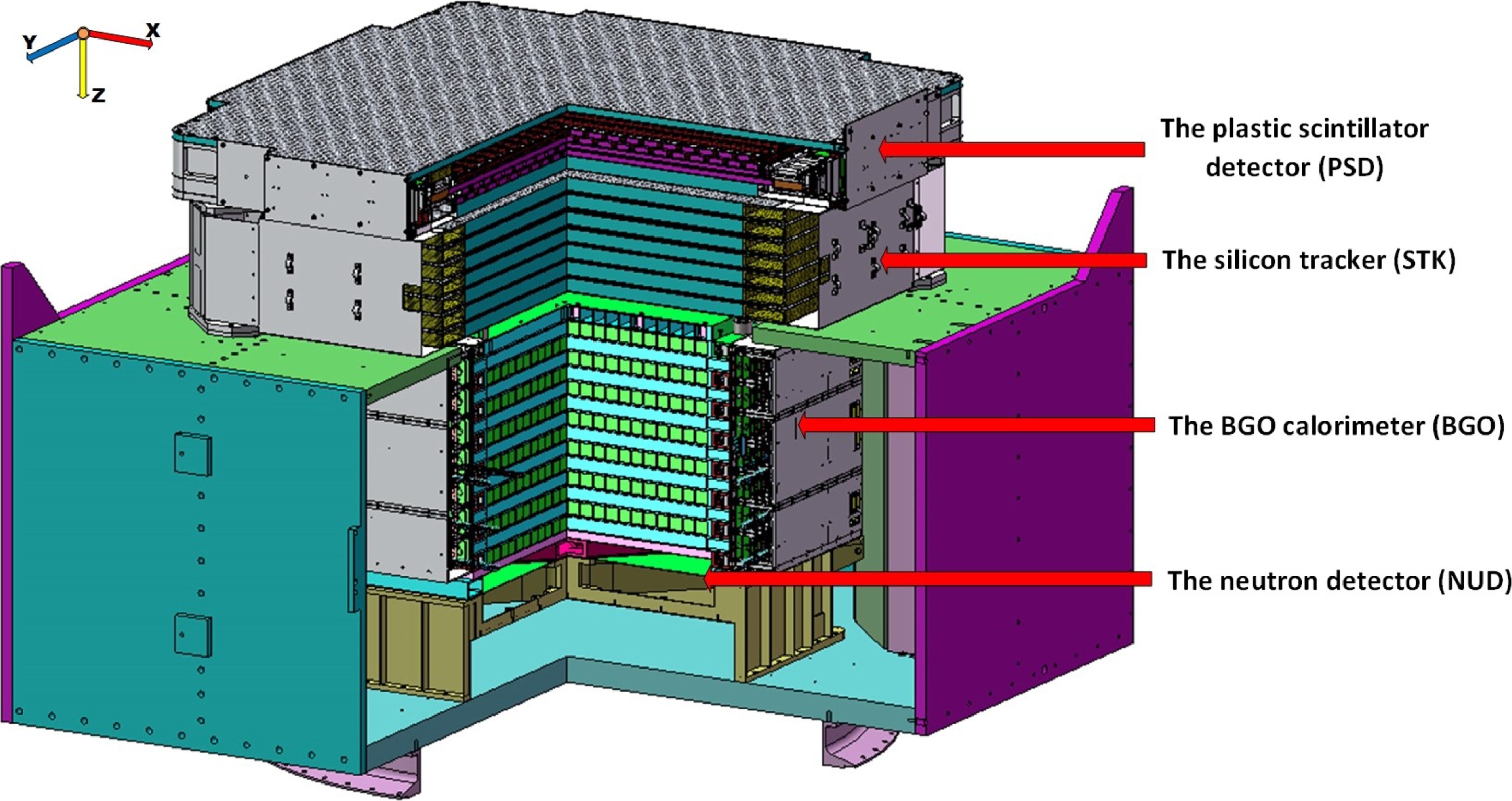}
\caption{Schematic view of DAMPE. It is composed of a plastic scintillator detector (PSD), a silicon–tungsten track (STK), a BGO calorimeter (BGO), and a neutron detector (NUD) \cite{CHANG20176}.\label{fig: DAMPE detector}}
\end{figure}

The Dark Matter Particle Explorer (DAMPE) is a Cosmic-Ray (CRs) and Gamma-Ray space-based observatory that was launched in 2015 \cite{CHANG20176}. The main goal of DAMPE is to investigate Cosmic-Ray Electrons plus Positrons\footnote{We will refer by electrons to electrons and positrons} (CREs) and Gamma-Rays up to 10 TeV; protons and heavier ions up to a few hundred TeV. The experiment is composed of 4 sub-detectors: a plastic scintillator (PSD) allowing the measurement of the charge\cite{YU20171}; a silicon-tungsten tracker-converter (STK) for precise measurement of the charge, track reconstruction and photon conversion \cite{AZZARELLO2016378}; a bismuth germaniun oxide (BGO) calorimeter \cite{WEI2019177} and a neutron detector (NUD) \cite{Huang_2020} (Figure~\ref{fig: DAMPE detector}).
\begin{figure}%[htbp]
\centering
\includegraphics[width=.7\textwidth]{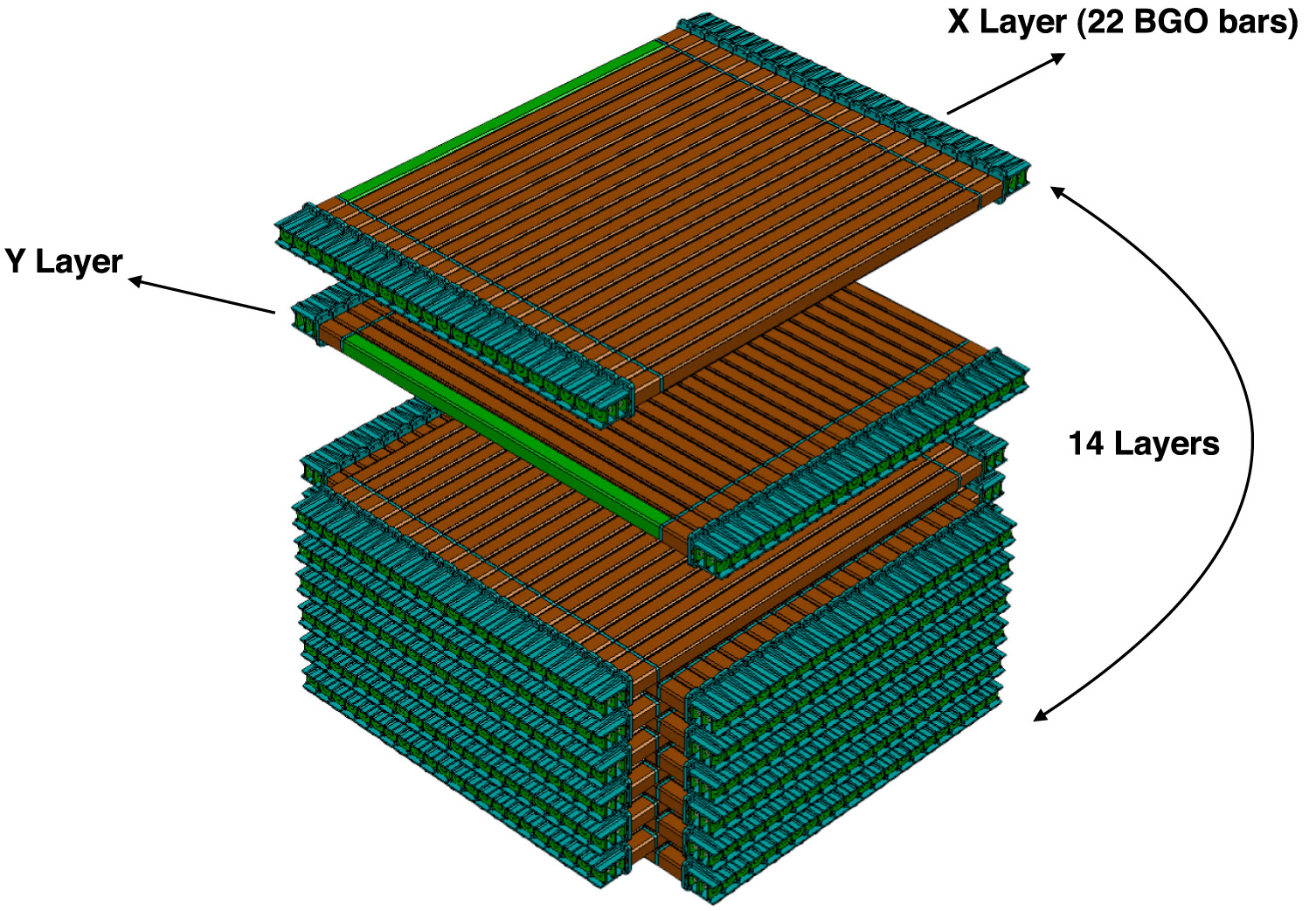}
\caption{View of the BGO-crystal bar arrangement in the calorimeter. There are 14 layers in a hodoscopic arrangement, each is composed of 22 bars \cite{WEI2019177}. \label{fig: BGO structure}}
\end{figure}These sub-detectors enable precise measurement of the absolute charge, energy and direction. The calorimeter is composed of 14 layers, each layer containing 22 BGO crystal bars, as shown in Figure~\ref{fig: BGO structure}, with an individual size of $25\times25\times600$ mm$^3$.

At high energies, CREs undergo energy losses during propagation due to their low mass, primarily from processes such as synchrotron radiation and inverse Compton scattering \cite{PhysRev.137.B1306}. Consequently, it is anticipated that electrons at energies higher than a few TeV originate from nearby sources, i.e. within a distance of about one kiloparsec, not older than about  $10^5$ years \cite{CREs_origin,2004ApJ...601..340K}. In addition, the aforementioned energy losses lead to a steeper energy spectrum which make them relatively rare at high energies, resulting on a lower CRE flux with respect to other species. Moreover, several dark matter models predict  the emission of electrons and positrons as a result of annihilation or decay \cite{DM_model,Conrad2017,PhysRevLett.122.041102}, which could then be seen as e.g. an excess in the high energy CRE spectrum \cite{PhysRevD.79.041301}. In 2017, DAMPE published its first measurement of the CRE flux using 1.5 years of data and featuring the direct detection of a break at 0.9 TeV \cite{DAMPE_elec_2017}. The energy reach of this measurement was limited to 5 TeV due to the low CRE flux (only 11 events were detected between 3 and 5 TeV). In order to enhance and expand our current understanding in the multi-TeV range, it is essential to increase the statistics and improve the accuracy. Hence, we enlarge the acceptance by selecting events outside of the fiducial volume, that we will call non-fiducial events, which are typically events with a large incidence angle. Since such events are less contained, a non-negligible fraction of their shower will develop outside of the detector. Consequently, the current energy reconstruction methods \cite{CHANG20176,BGO_correction}, which are typically applied to events nearly perpendicular to the detector, become less efficient at accurately determining the particle's energy.

In this paper we propose a new method using Convolutional Neural Networks (CNNs) for the energy reconstruction of electrons on the DAMPE experiment. In section \ref{section:Non-fiducial Events}, we introduce the concept of non-fiducial events. Section \ref{section:Classical Energy Reconstruction of Non-fiducial events} presents the classical method and the reasoning for transitioning to a neural network-based approach. Section \ref{section:Convolutional method to correct the reconstructed energy} showcases the CNN model and training. In section \ref{section:Results}, we present the results and compared the performances of this methods with the classical one. Additionally, we demonstrate how the CNNs enables to avoid the spectral unfolding. In section \ref{section:Flight data comparison}, we show the application of the CNN to data and compare it to Monte-Carlo (MC) simulations. Finally, in section \ref{section: Conclusion} we summarize the findings and contributions of this work.

\section{Non-fiducial Events}
\label{section:Non-fiducial Events}
\begin{figure}%[htbp]
\centering
\quad
\includegraphics[width=.7\textwidth]{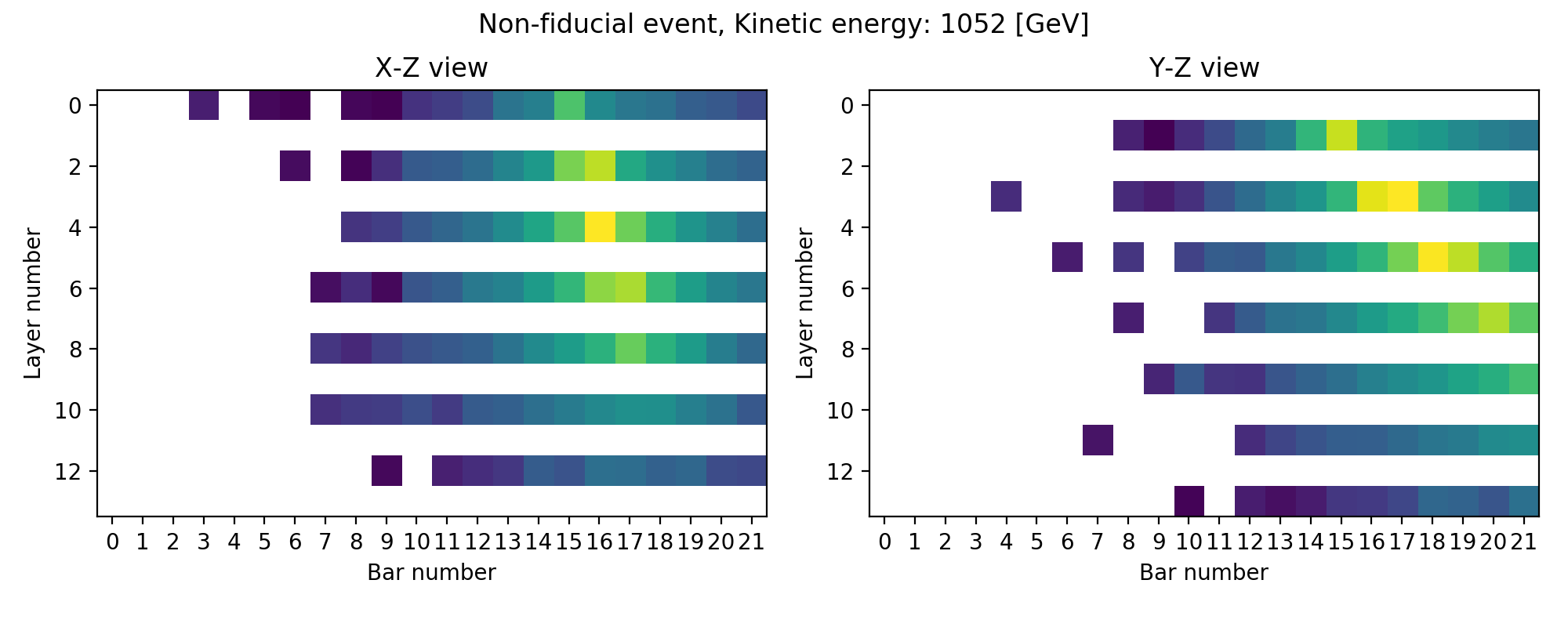}
\qquad
\includegraphics[width=.7\textwidth]{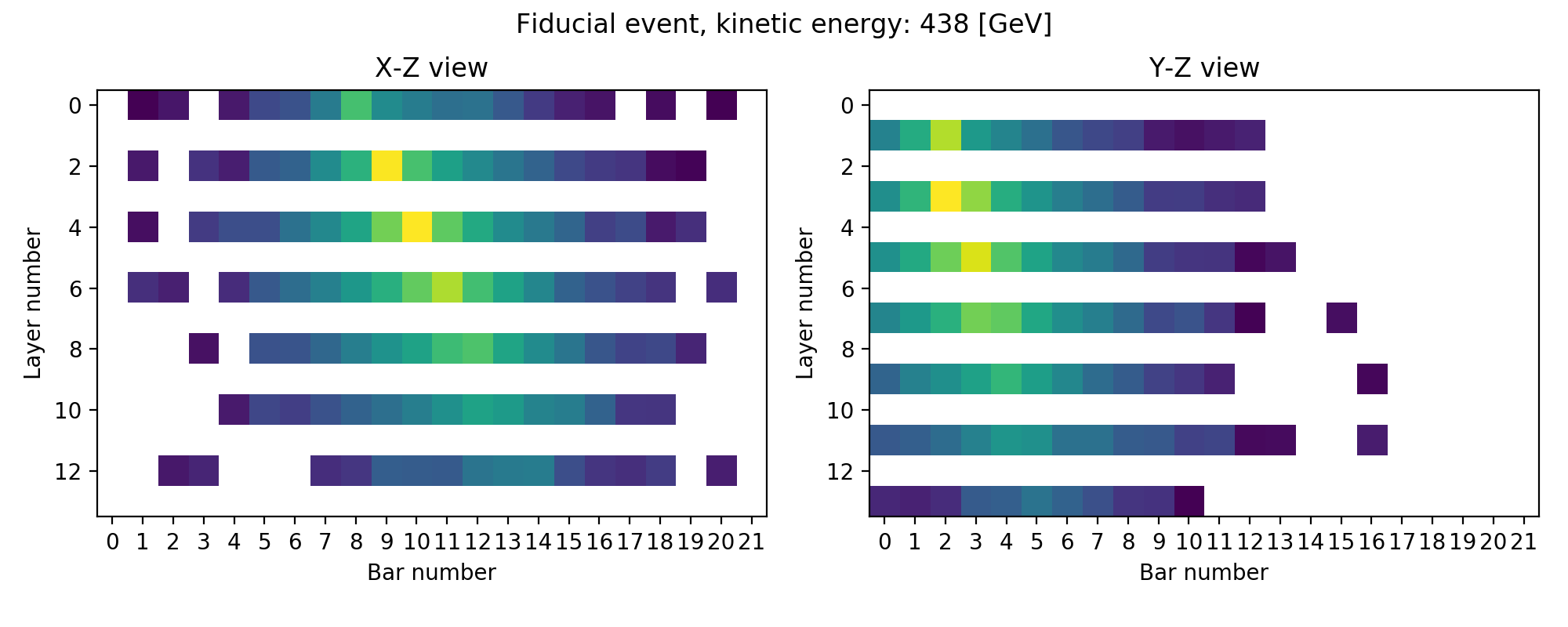}
\caption{View of a non-fiducial event (top figure) and fiducial event (bottom figure) crossing the calorimeter. Left plots show the X-Z view while right plots show the Y-Z views.\label{fig: Fid vs NonFid}}
\end{figure}

To motivate the cuts of the non-fiducial event selection, we start by describing the default fiducial selection used in previous publications \cite{DAMPE_elec_2017,Droz_ML}. We start by following a set of initial selections based on the shower development in the calorimeter, defined as follows:

\begin{itemize}
    \item Rejecting events that enter the calorimeter from the sides, by imposing a energy fraction in one layer to not exceed $35\%$. Specifically, we identify the bar with the highest energy deposition and, if the ratio of this energy to the total deposited energy exceeds 0.35, the event is excluded.
    \item Rejecting events where the shower direction cannot be reconstructed.
    \item Ensuring that the reconstructed shower direction extrapolates to the top and bottom of the BGO sensitive volume, within a distance of 280 mm from the centre, in either the X or Y direction.
\end{itemize}
These cuts aim to select events that are well reconstructed and contained in the calorimeter.
We define non-fiducial events (Figure~\ref{fig: Fid vs NonFid}) by inverting the last cut, i.e. by requiring that the top or bottom extrapolation of the shower is not within a distance of 280 mm from the centre. By inverting only the third cut, we can retain events that have a signal in the other detectors, which are used for flux measurements.

On top of that, the analysis chain involves standard preselection cuts to reject ions, obvious protons, or poorly reconstructed events in a similar fashion to the 2017 analysis \cite{DAMPE_elec_2017}. 

\section{Classical Energy Reconstruction of Non-fiducial events}
\label{section:Classical Energy Reconstruction of Non-fiducial events}
The standard method for inferring particle energy \cite{BGO_correction} relies on the total energy deposition in the calorimeter. However, this technique was primarily developed for fiducial events,which are typically associated with low incidence angles. When applied to non-fiducial events, the performance degrades because the shower development occurs largely in the dead material, such as the carbon-fibres and rubbers used as the support structure, and outside the calorimeter, leading to poorer reconstruction and missing energy.

As illustrated in Figure~\ref{fig: Fid vs NonFid}, non-fiducial events often have complex shower topologies. This complexity requires the use of more sophisticated techniques for accurate energy reconstruction. Classical methods relied on the deposited energy in each BGO bar, which was obtained from the multi-dynode readout. These techniques enable reconstructing energies up to 10 TeV with a resolution of $1.2 \%$ at energies above 100 GeV for electrons/positrons \cite{CHANG20176}. Generally, the reconstructed energy will be lower than the true energy due to losses in the dead material region between the bars as well as the adjacent bars and the front of the full detector. The work in \cite{BGO_correction} developed a method to correct for the missing energy. Nevertheless, both methods were optimized for fiducial electrons, for which the lateral and longitudinal shower development is well understood. To evaluate the efficiency of various reconstruction methods, we compare them by examining the relative error, defined by:
\begin{equation}
    r = 1- \frac{E}{E_\mathrm{kin}},
\end{equation}
where $E$ is the reconstructed energy and $E_\mathrm{kin}$ is the initial energy of the particle.

\begin{figure}[htbp]
    \centering
    % First row of images
    \begin{subfigure}[b]{0.45\textwidth}
        \centering
        \includegraphics[scale=0.45]{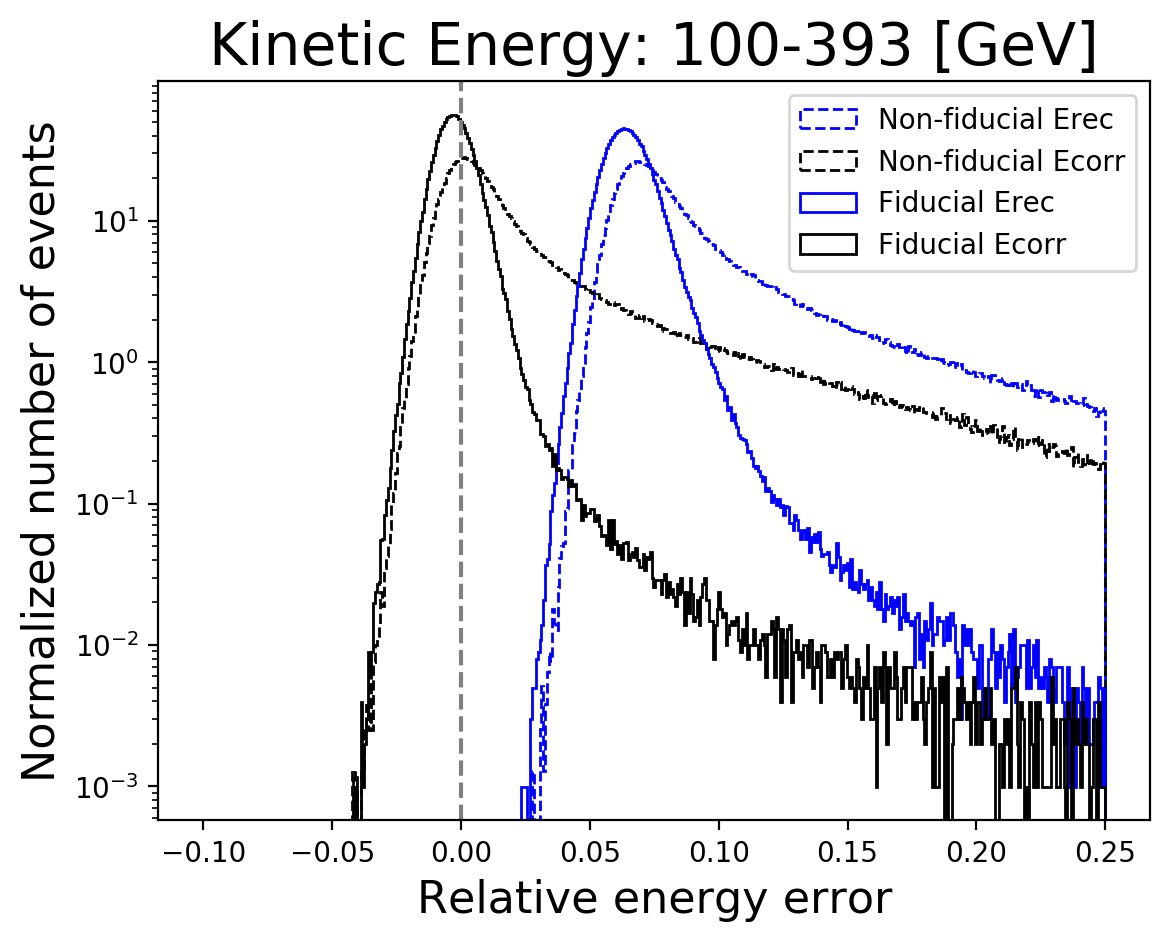}
        \caption{100-393 GeV}
    \end{subfigure}
    \hfill
    \begin{subfigure}[b]{0.45\textwidth}
        \centering
        \includegraphics[scale=0.45]{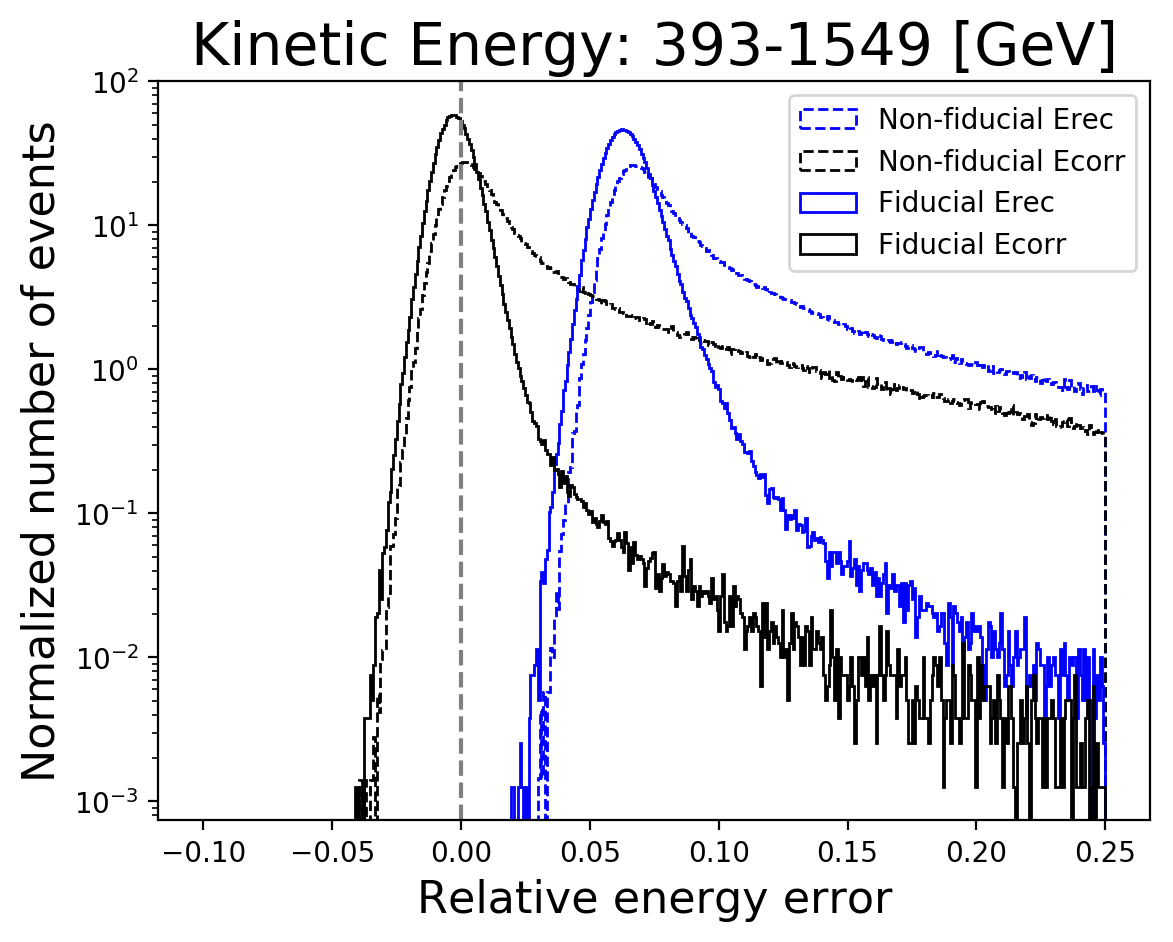}
        \caption{393-1549 GeV}
    \end{subfigure}
    
    \vskip\baselineskip % Space between rows

    % Second row of images
    \begin{subfigure}[b]{0.45\textwidth}
        \centering
        \includegraphics[scale=0.45]{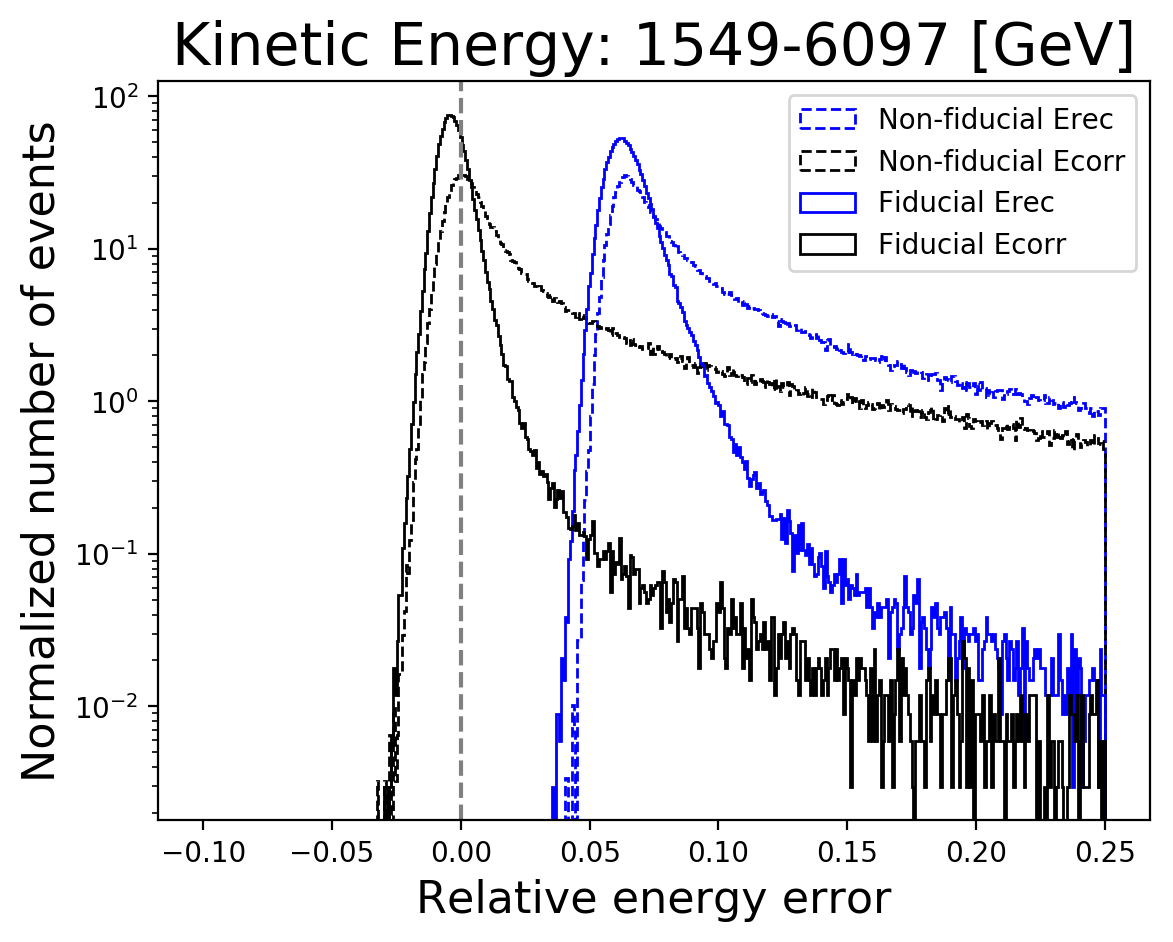}
        \caption{1549-6097 GeV}
    \end{subfigure}
    \hfill
    \begin{subfigure}[b]{0.45\textwidth}
        \centering
        \includegraphics[scale=0.45]{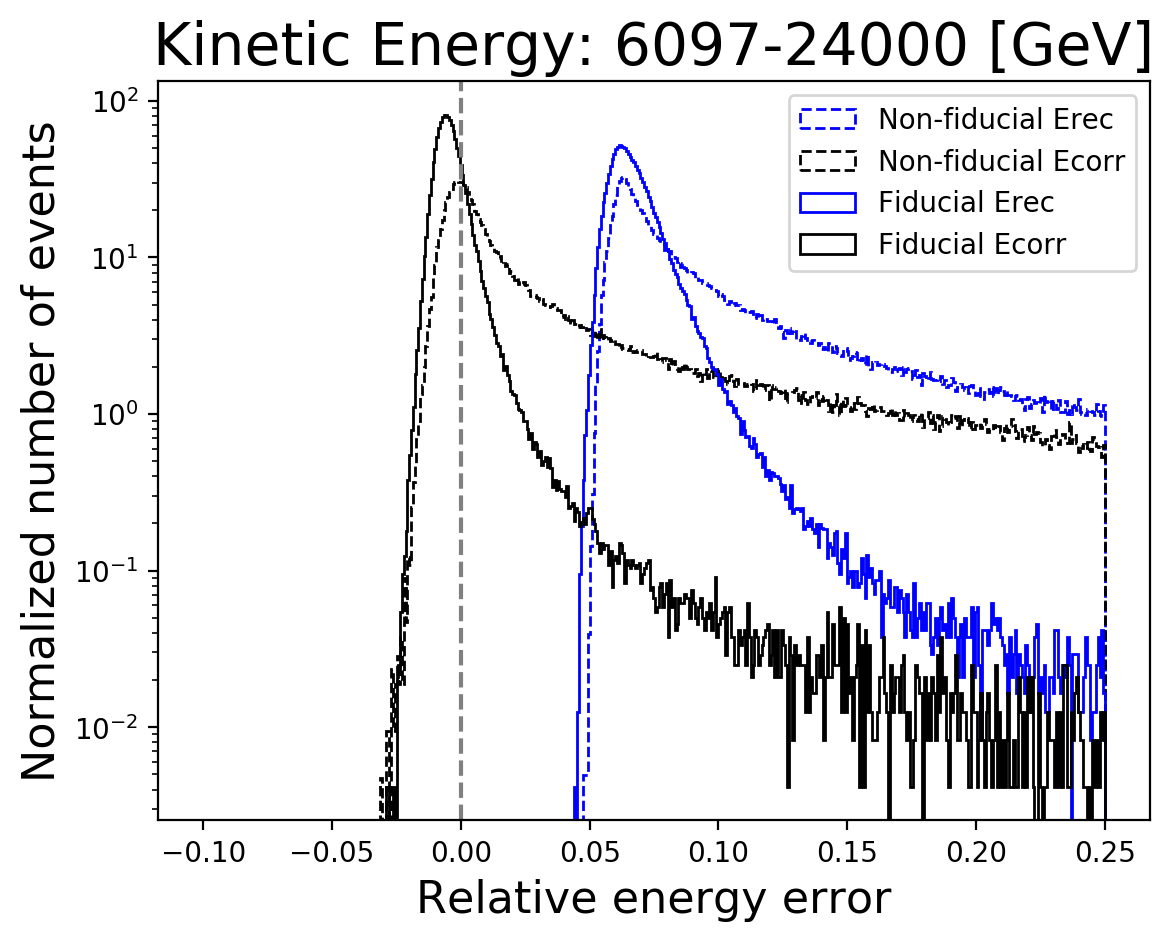}
        \caption{6097-24000 GeV}
    \end{subfigure}
    
    \caption{Distribution of the relative energy error obtained with electron MC simulations in four different energy bins, ranging from 100 GeV to 24 TeV. Two types of MC samples are used -- fiducial and non-fiducial, analysed with two energy reconstruction methods -- standard \cite{CHANG20176} and corrected \cite{BGO_correction}. Distributions are denoted as follows: fiducial standard (blue full line), non-fiducial standard (blue dashed line), fiducial corrected (black full line) and non-fiducial corrected (black dashed line). The grey line indicates ideal case when reconstructed energy equals the true kinetic energy.}
    \label{fig: ratio Erec and Ecorr}
\end{figure}

In Figure~\ref{fig: ratio Erec and Ecorr}, we show the comparisons of methods from \cite{CHANG20176} and \cite{BGO_correction}. As can be seen, the energy reconstructed with classical methods $E_\mathrm{rec}$ is systematically underestimated by at least 5\% compared to the kinetic energy. As mentioned earlier, this is expected due to shower leakage and energy losses in the dead material. However, the corrected energy evaluation, $E_\mathrm{corr}$, developed in \cite{BGO_correction}, accounts for those effects.

\begin{figure}%[htbp]
\centering
\includegraphics[width=.6\textwidth]{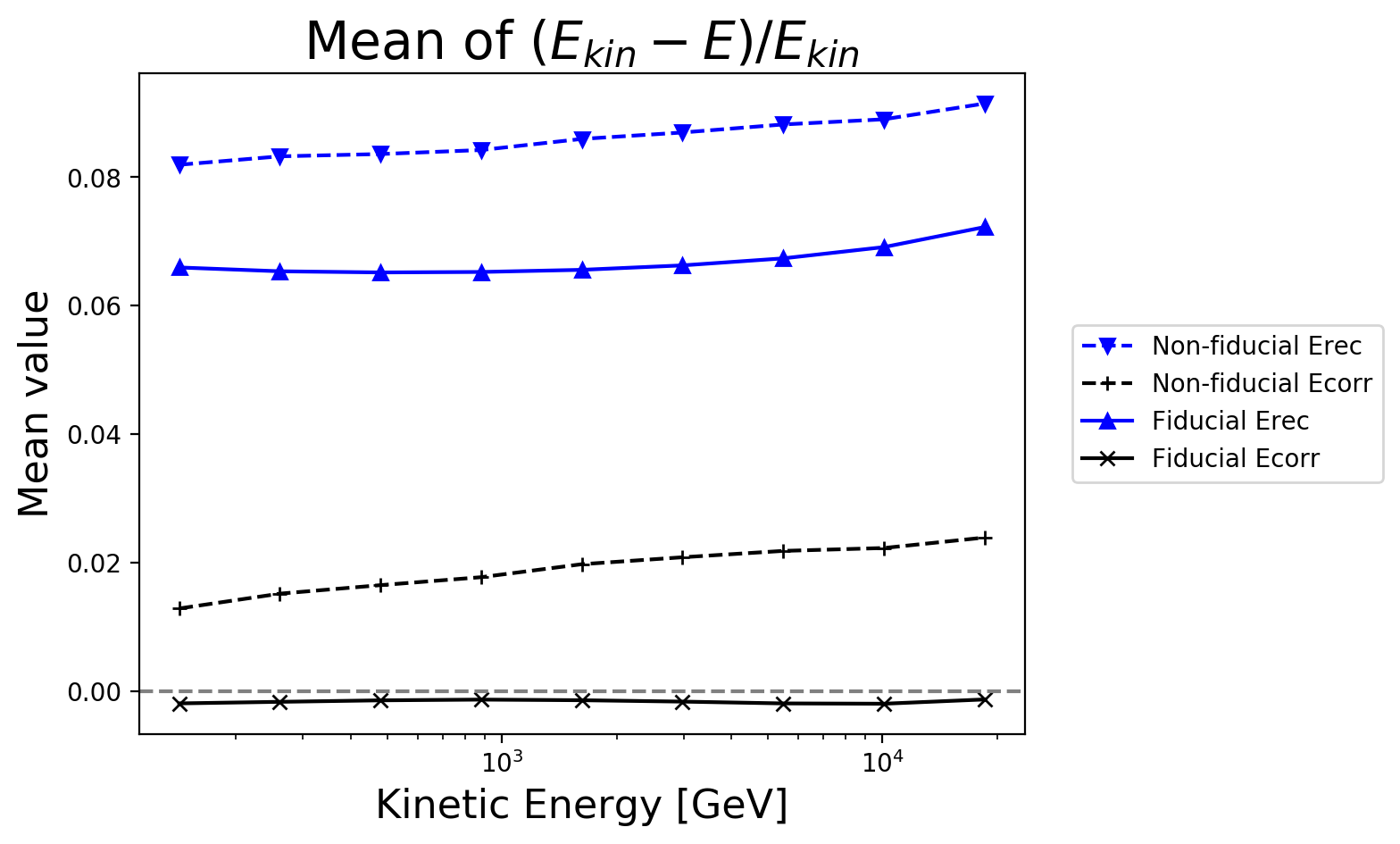}
\caption{Mean of the relative energy error as a function of kinetic energy for the different energy reconstruction methods.\label{fig: mean ratio}}
\end{figure}

\begin{figure}%[htbp]
\centering
\includegraphics[width=.5\textwidth]{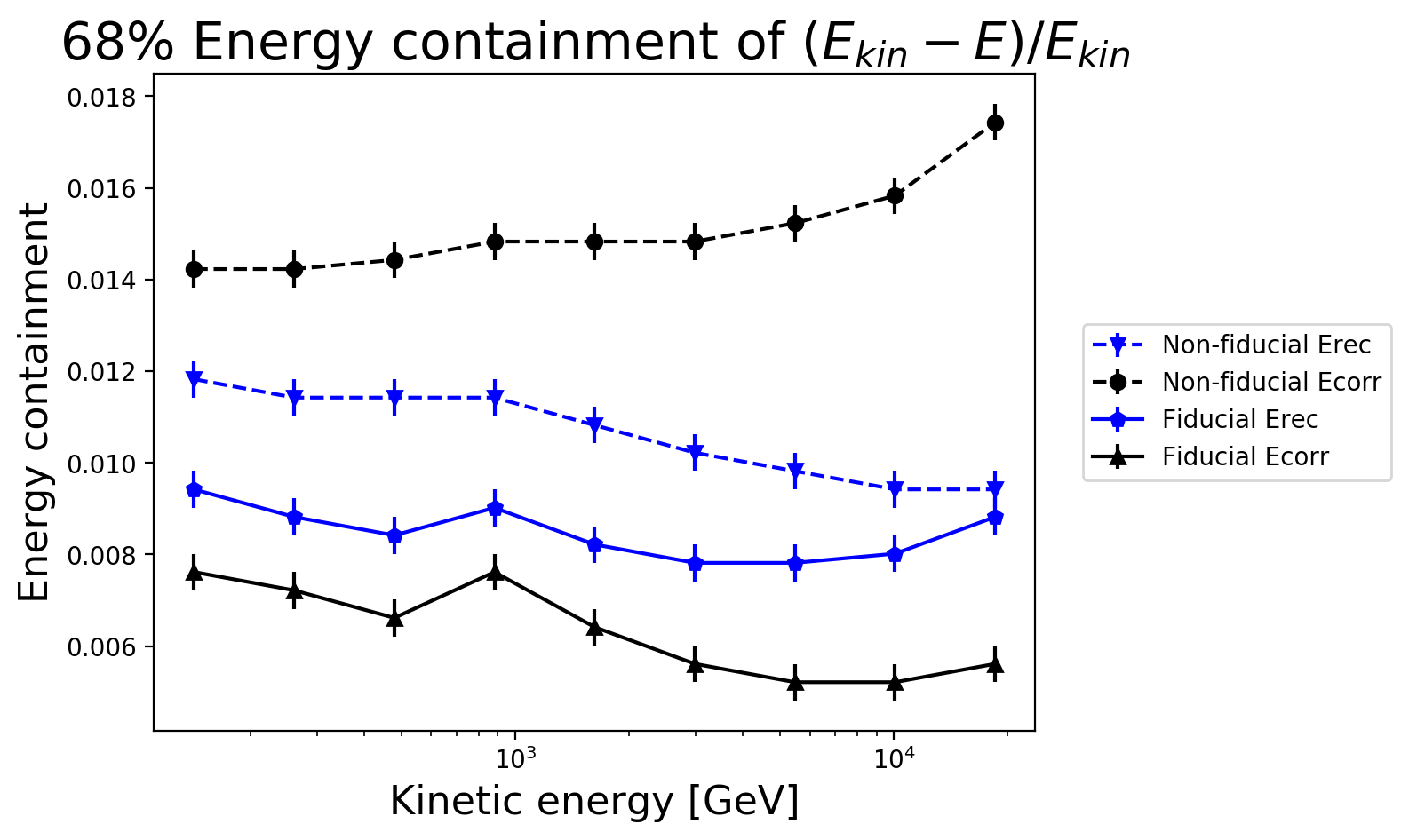}
\qquad
\includegraphics[width=.5\textwidth]{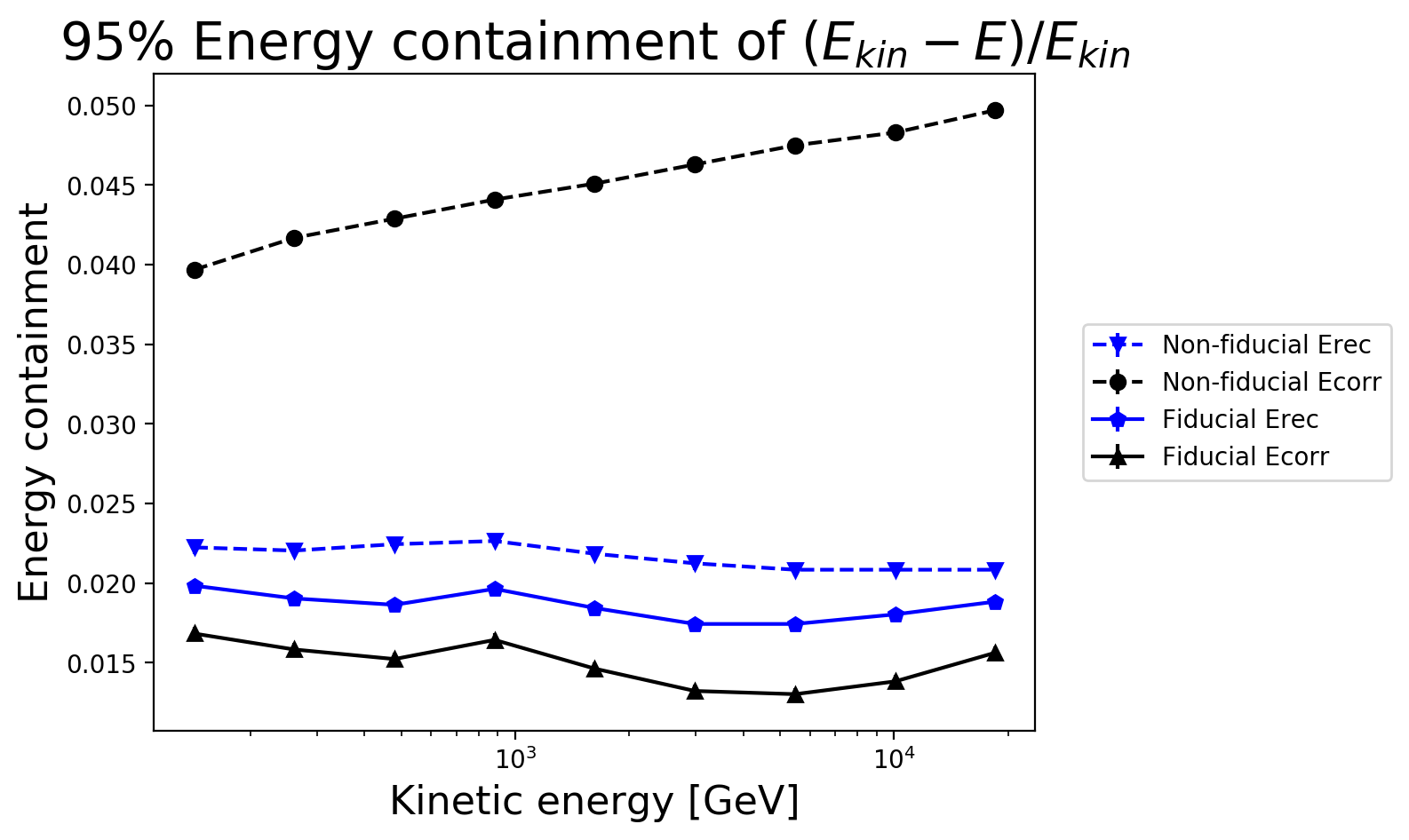}
\caption{Energy resolution as a function of kinetic energy. Top plot: $68\%$ containment, bottom plot: $95\%$ containment. \label{fig: radius ratio}}
\end{figure}

 The corrected method ($E_\mathrm{corr}$) \cite{BGO_correction} provides a better estimation of the true particle energy, as the bulk of the distribution has an estimated energy closer to the kinetic energy, as shown in Figures~\ref{fig: ratio Erec and Ecorr} and \ref{fig: mean ratio}. However, there is a pronounced tail towards the positive error values, similar to the one of $E_\mathrm{rec}$, which is due to events with large energy leakage that cannot be well contained by the calorimeter. On top of that, the energy resolution for both methods becomes worse with increasing energy, as shown in Figure~\ref{fig: radius ratio}. 

Overall, we observe that $E_\mathrm{corr}$ is optimized for fiducial events. To correct the energy of non-fiducial events, a new method is required. It is developed in this paper using CNNs approach, where a neural network is train to predict the relative error of the true energy and the reconstructed energy. The CNN will be presented in the next section, where we will also evaluate its accuracy.

\section{Convolutional method to correct the reconstructed energy}
\label{section:Convolutional method to correct the reconstructed energy}
Machine Learning (ML) techniques have been widely applied in various domains of physics \cite{ML_allPhysics}. The DAMPE collaboration has utilized such techniques for tasks ranging from the discrimination between protons and electrons \cite{Droz_ML,PCA}, to track reconstruction \cite{Andrii_ML}, and calorimeter saturation correction \cite{Misha_ML}.

To accurately recover the true kinetic energy of non-fiducial events, we train a CNN to perform a regression of the relative error between the reconstructed energy ($E_\mathrm{rec}$) and the true energy of the particle ($E_\mathrm{kin}$), defined by:
\begin{equation}
    r = 1- \frac{E_\mathrm{rec}}{E_\mathrm{kin}}.
    \label{Eq: ratio def}
\end{equation}

We chose to use the relative error instead of the true kinetic energy because it allows the model to achieve better performances, as the values of r are bounded between $0$ and $1$. The goal is to obtain a prediction for the ratio that will  then be used to correct the energy.
As an input to the CNN we use an image of the BGO calorimeter, combining the X-Z and Y-Z views, as illustrated in Figure~\ref{fig: Fid vs NonFid}. Since the layers of the calorimeter are placed in a hodoscopic arrangement, this results in a correlation between both views. As a result, we expect that training a CNN with each view separately would result in a poorer resolution, as discussed in \cite{Andrii_ML,Misha_ML} since both views are correlated. The input to the CNN is normalise by the maximum deposited energy in the calorimeter resulting in images with pixel values between 0 and 1.

The model was trained based on MC simulations of electrons. The training was done based on Monte-Carlo (MC) simulations, generated using Geant4 (version 10.05) \cite{AGOSTINELLI2003250} with FTFP-BERT physics. The simulation samples are weighted following a $E^{-3}$ power law. The MC samples were processed following the selection procedure described in section~\ref{section:Non-fiducial Events}. We divide MC into a training sample and a validation sample by selecting $70\%$ of the full MC sample for the training and the remaining $30\%$ as validation sample. To be able to have an uniform performance over all energies, we split the sample into 10 bins equally spaced in logarithm of kinetic energy (from 100 GeV to 24 TeV), then choose the energy bin with the lowest number of events and select the same number of events in each of the other energy bins. The resulting training sample was composed of $1'084'912$ events.

We used Tensorflow 2 \cite{tensorflow2015-whitepaper} as the framework for training of the CNN. The model structure is the following one:
\begin{figure}[htbp]
\centering
\includegraphics[scale=0.4]{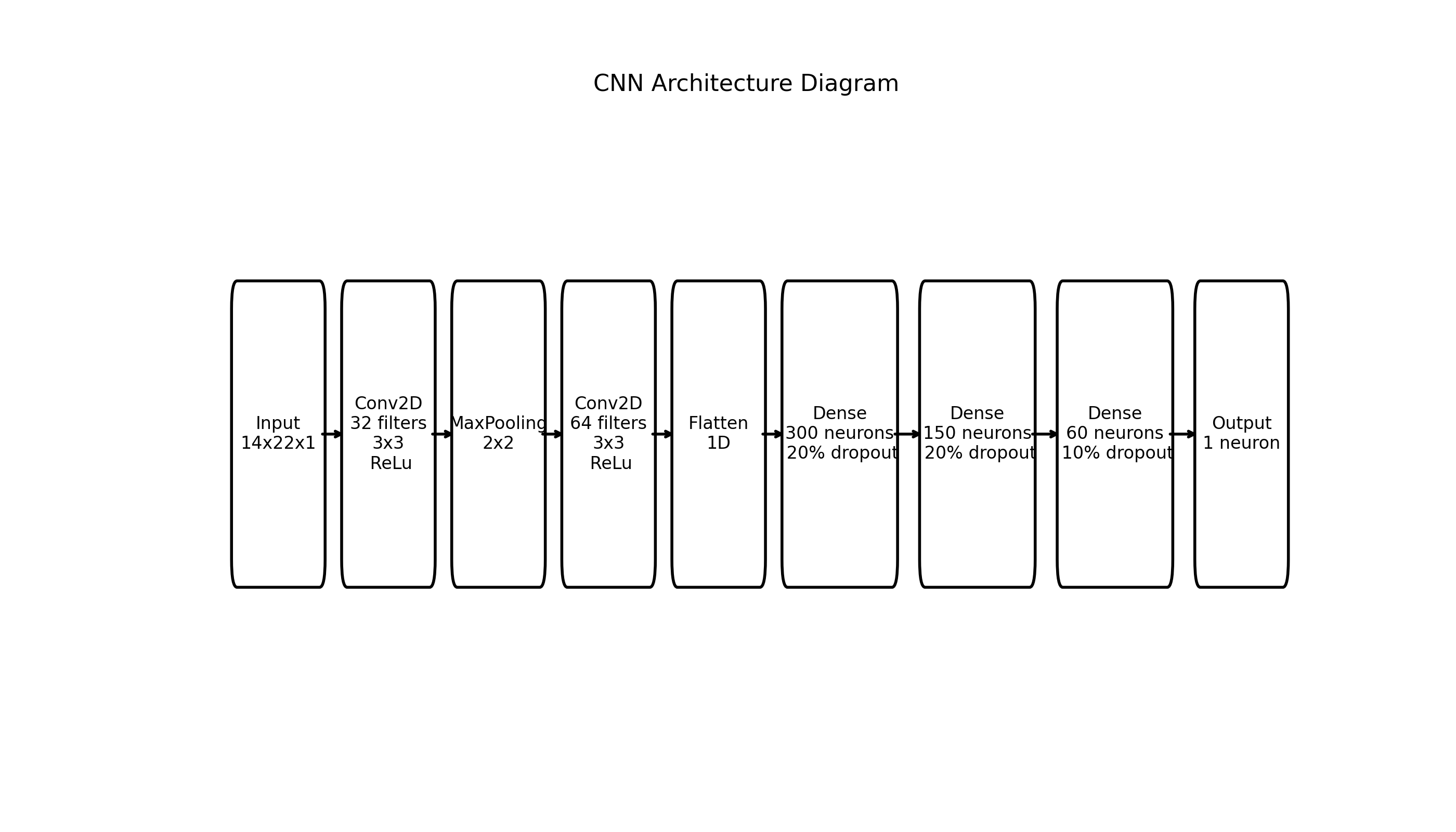}
\caption{CNN model structure.\label{fig: model structure}}
\end{figure}

The convolutional layers in the proposed CNN utilise 3×3 kernels. This kernel size is a common choice in deep learning models, including well-known architectures such as VGGNet \cite{Simonyan15}, due to its favourable trade-off between spatial feature extraction and computational efficiency. A 3×3 kernel effectively captures local patterns (e.g., edges, corners, and textures) while maintaining a manageable number of parameters. 
The activation function used is a ReLU. It has become the default activation function for many types of neural networks \cite{NIPS2012_c399862d, Andrii_ML, Misha_ML} because a model that uses it is easier to train and often achieves better performance. Finally, as for loss function, we choose the mean squared error (MSE) which is a common loss function used in regression tasks. It measures the average squared difference between the predicted value and the true values, ensuring that larger errors are penalised more heavily than smaller ones. To regularise the loss we added dropout layers, the key idea is to randomly drop units (along with their connections) from the neuron part of the model. This prevents the model to overfit during training. I also encourages the network to not rely too heavily on any single feature or neuron, leading to more robust results against small changes in the weights or training data. 
Finally, as a loss function to evaluate whether the CNN training goes under- or over- fitting we look at the evolution of the mean squared error as a function of the epochs, defined by:

\begin{equation}
    \frac{1}{\mathrm{m}}\sum_{i=0}^{\mathrm{m}}(\hat{r}_{i}-r_{i})^2,
    \label{Eq: mean squared error}
\end{equation}
where m is the number of training samples, $r_{i}$ is the relative error between $E_\mathrm{rec}$ and $E_\mathrm{kin}$ as defined in Equation~\ref{Eq: ratio def}, and $\hat{r}_i$ is the CNN prediction.

\begin{figure}[htbp]
\centering
\includegraphics[width=.75\textwidth]{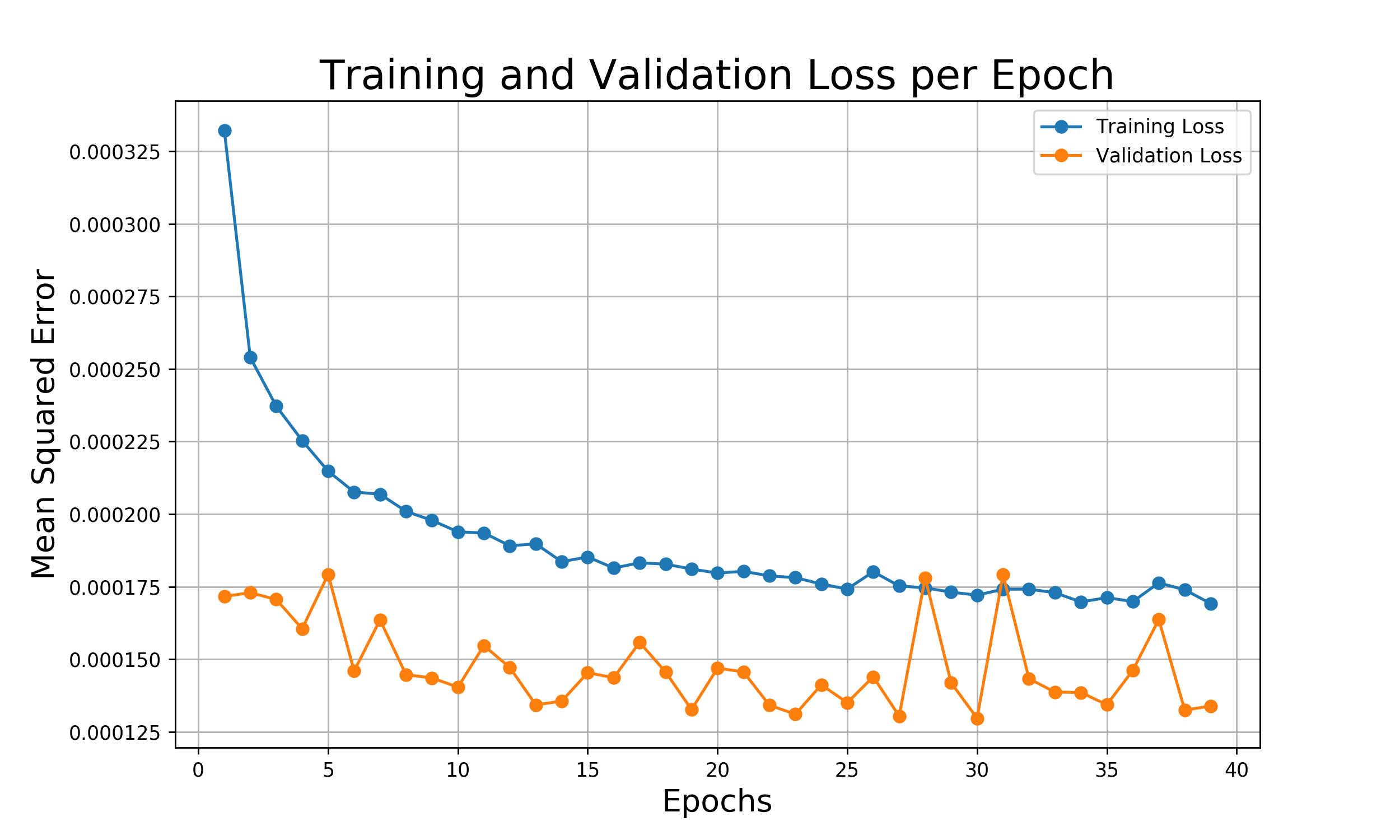}
\caption{Evolution of the loss, defined as the mean squared error for the training sample (blue curve) and validation sample (orange curve).\label{fig: CNN loss}}
\end{figure}

Figure~\ref{fig: CNN loss} shows the convergence of the training and validation sample after 40 epochs. As we can see the loss has converged close to $1.75\cdot 10^{-4}$ ($1.50\cdot 10^{-4}$ for the validation loss). While the loss for the validation sample has greater oscillations, this can be explained by the fact that its composition is different at all energies, meaning that there is no balance between low energy and high energy.

\section{Results}
\label{section:Results}
The energy output is obtained from the CNN output, according to Equation \ref{Eq: ratio def}, as follows:
\begin{equation}
    E_\mathrm{CNN} = \frac{E_\mathrm{rec}}{1-r},
\end{equation}
%where r is the CNN output and $E_\mathrm{rec}$ is the reconstructed Energy.

To estimate the CNN's efficiency and compare it to the classical methods (see Figure~\ref{fig: ratio Erec and Ecorr}), we also calculate the relative error for $E_\mathrm{CNN}$, defined by:
\begin{equation}
    r_{\mathrm{CNN}}=1-\frac{E_\mathrm{CNN}}{E_\mathrm{kin}}.
\end{equation}

\begin{figure}[htbp]
\centering
\includegraphics[scale=0.4]{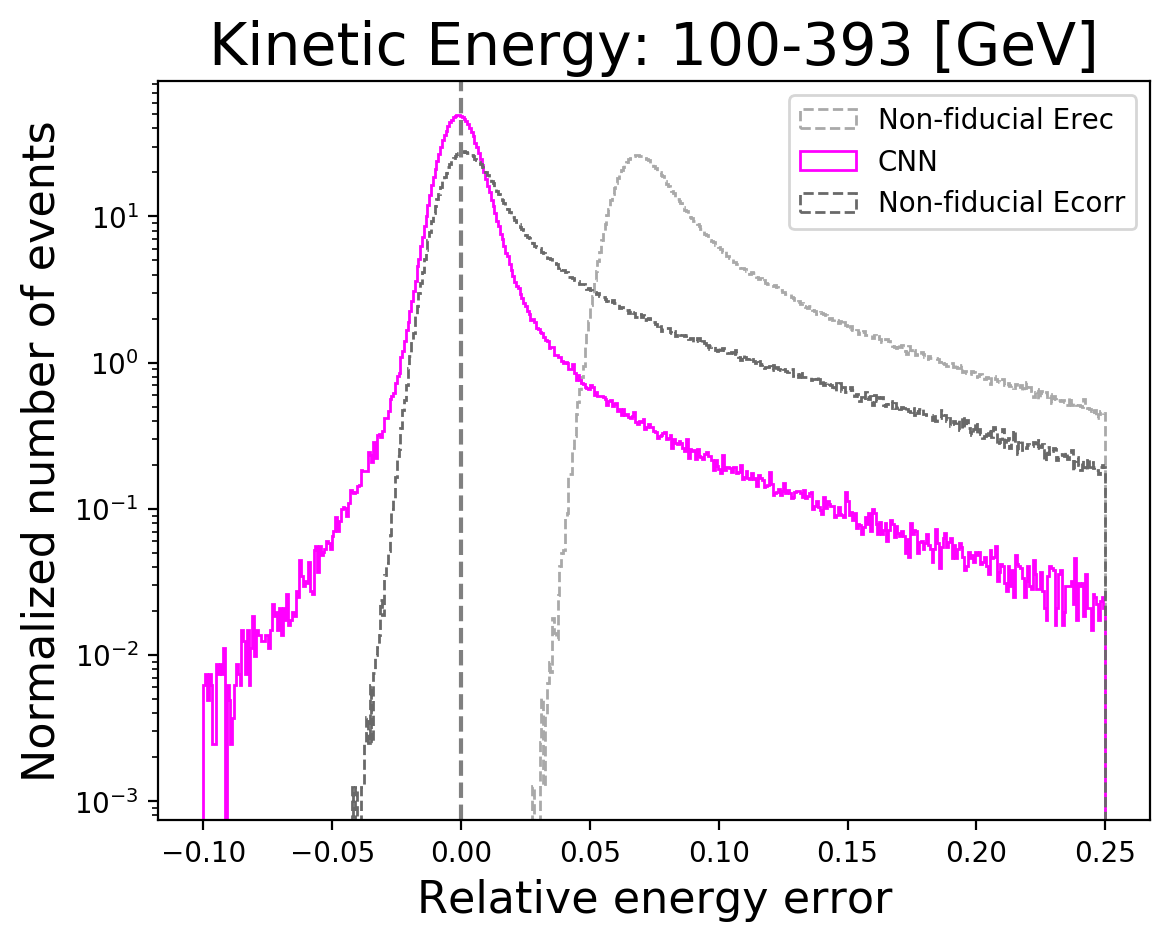}
\qquad
\includegraphics[scale=0.4]{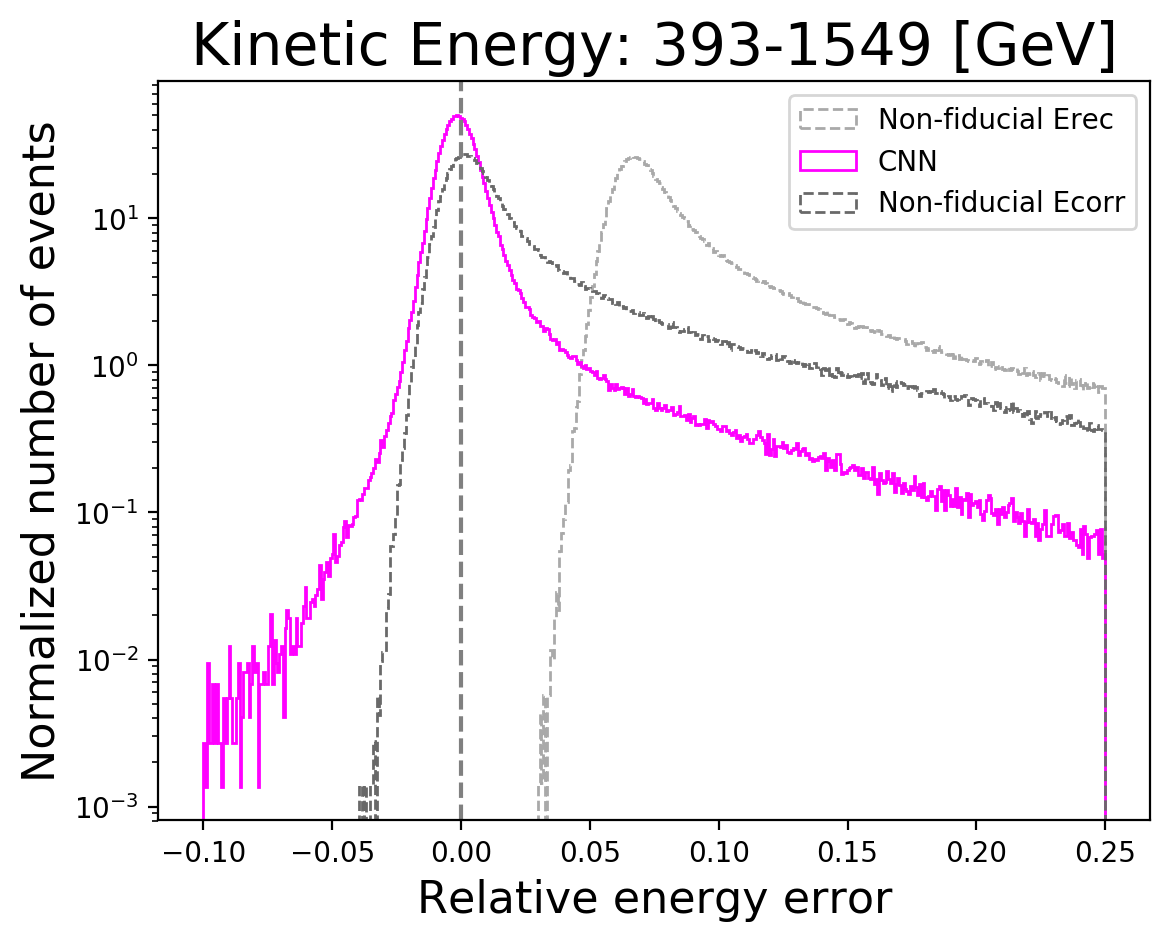}
\qquad
\includegraphics[scale=0.4]{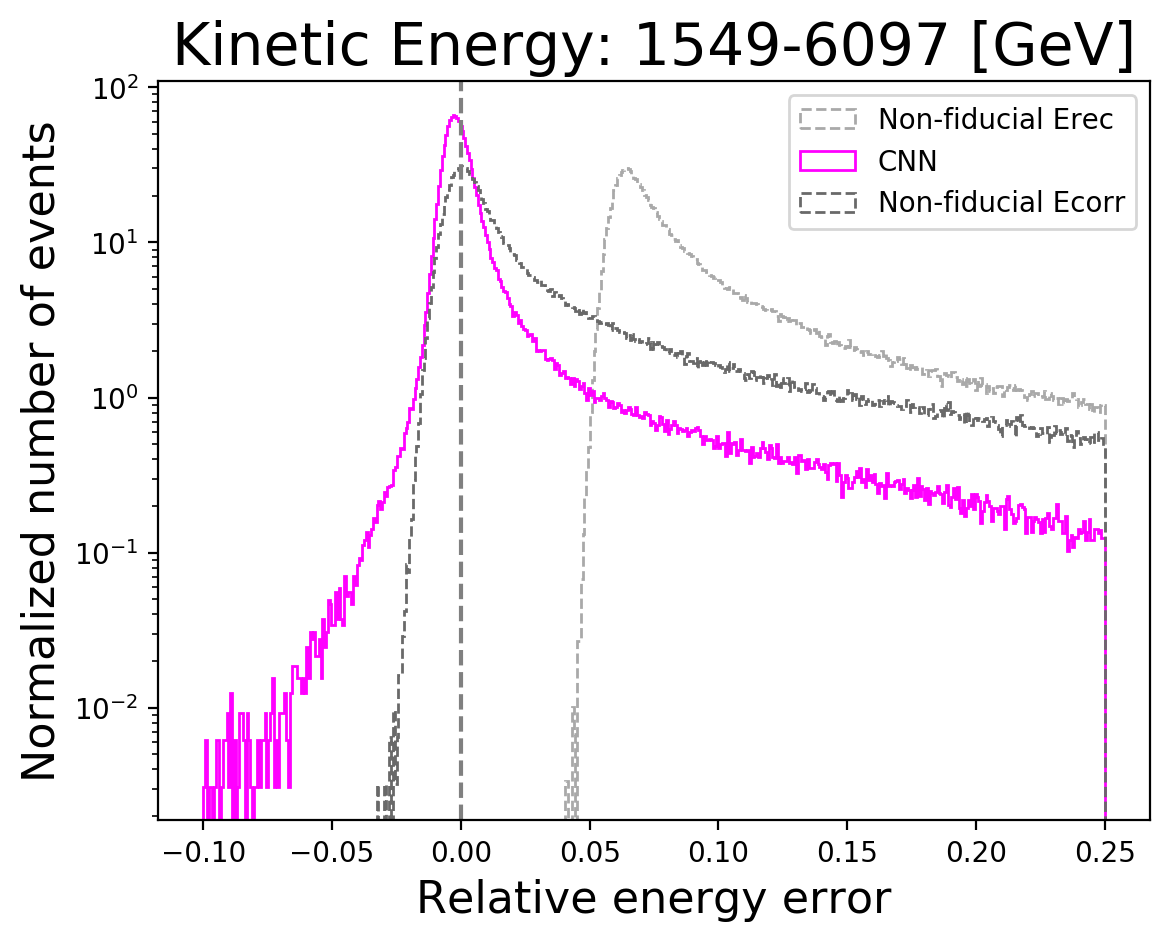}
\qquad
\includegraphics[scale=0.4]{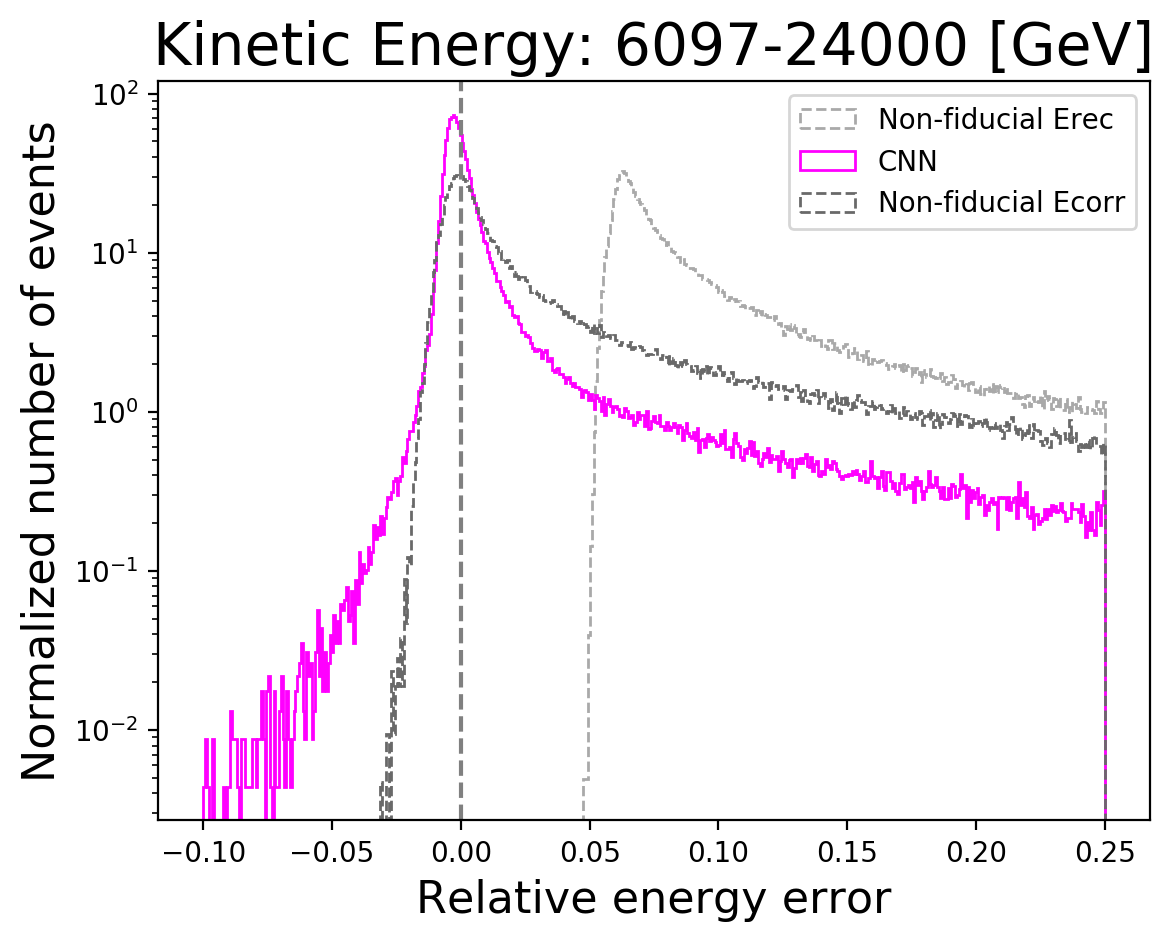}
\caption{The distribution of the relative energy error (Equation~\ref{Eq: ratio def}) for electron MC samples ranging from 100 GeV to 24 TeV. Light grey curve: for the standard energy reconstruction method \cite{CHANG20176}, dark grey curve: standard electron energy correction method \cite{BGO_correction} and magenta curve: CNN method. Grey vertical line indicates ideal case when reconstructed energy equals the true kinetic energy. \label{fig: ratio CNN}}
\end{figure}

In Figure~\ref{fig: ratio CNN}, we compare the reconstructed energy obtained by the classical methods (blue curve and yellow curve) \cite{BGO_correction} and the CNN (green curve) for electron MC. While the CNN was trained on electrons only, its performance is expected to be identical for both electrons and positrons since their electromagnetic showers are effectively indistinguishable at energies at energies sufficiently higher than 10 MeV (see Figure 34.11 from \cite[page 582]{ParticleDataGroup:2024cfk}).

\begin{figure}[htbp]
\centering
\includegraphics[width=.6\textwidth]{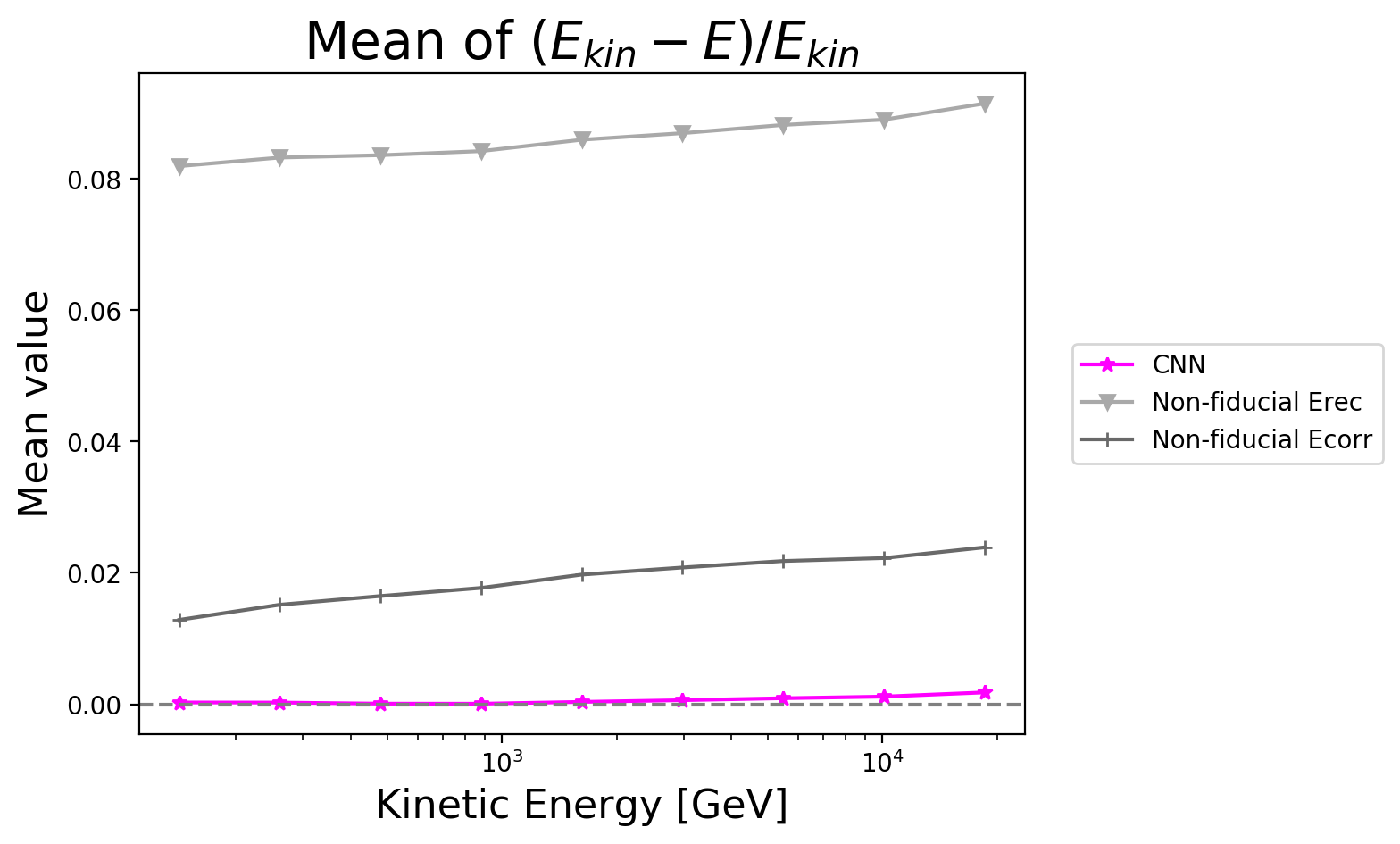}
\caption{Plot of the mean of the relative energy error (Equation~\ref{Eq: ratio def}) for electron MC samples ranging from 100 GeV to 24 TeV. Light grey curve: for the standard energy reconstruction method \cite{CHANG20176}, dark grey curve: standard electron energy correction method \cite{BGO_correction} and magenta curve: CNN method. Horizontal grey line indicates ideal case when reconstructed energy equals the true kinetic energy. \label{fig: mean ratio CNN}}
\end{figure}

\begin{figure}[htbp]
\centering
\includegraphics[width=.5\textwidth]{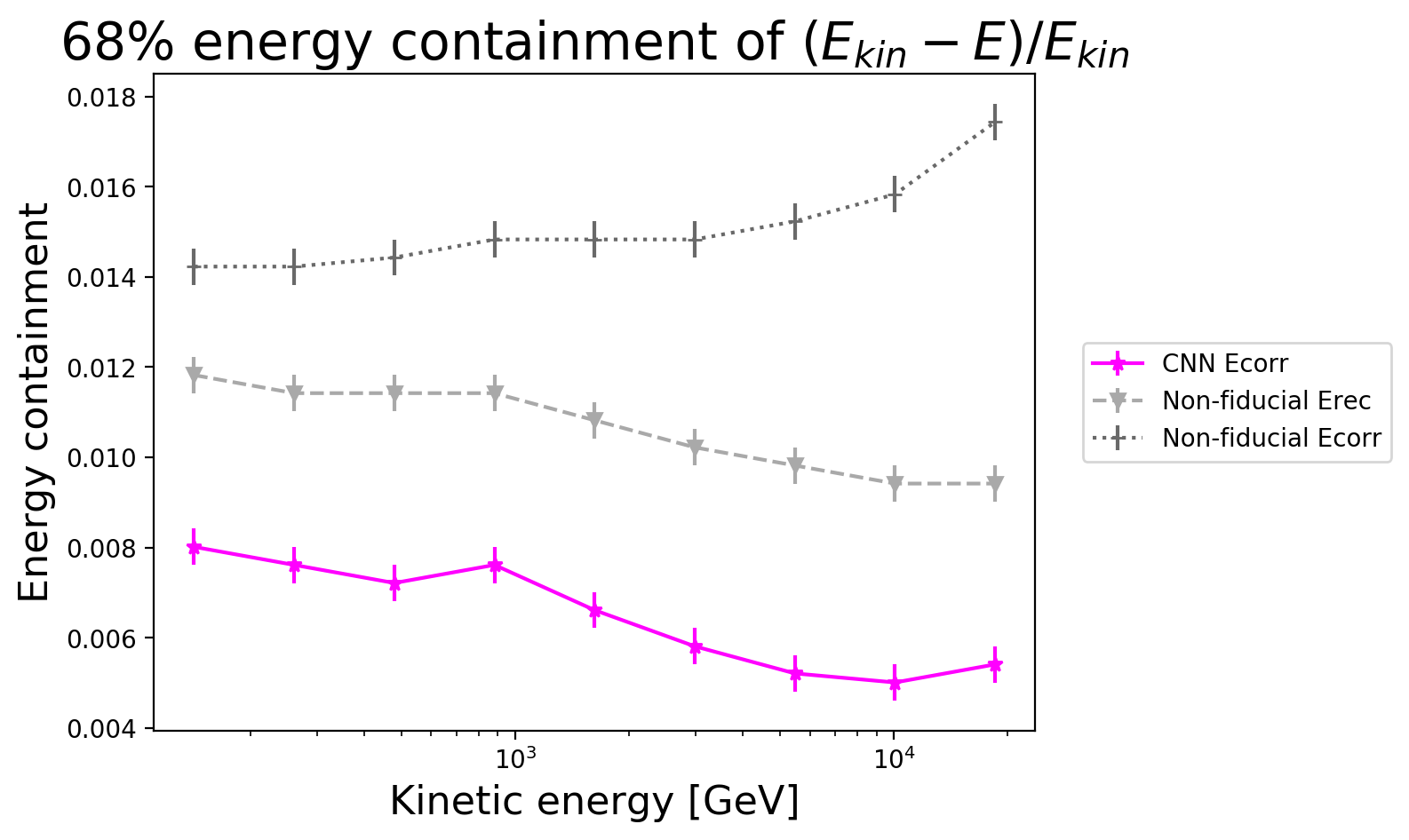}
\qquad
\includegraphics[width=.5\textwidth]{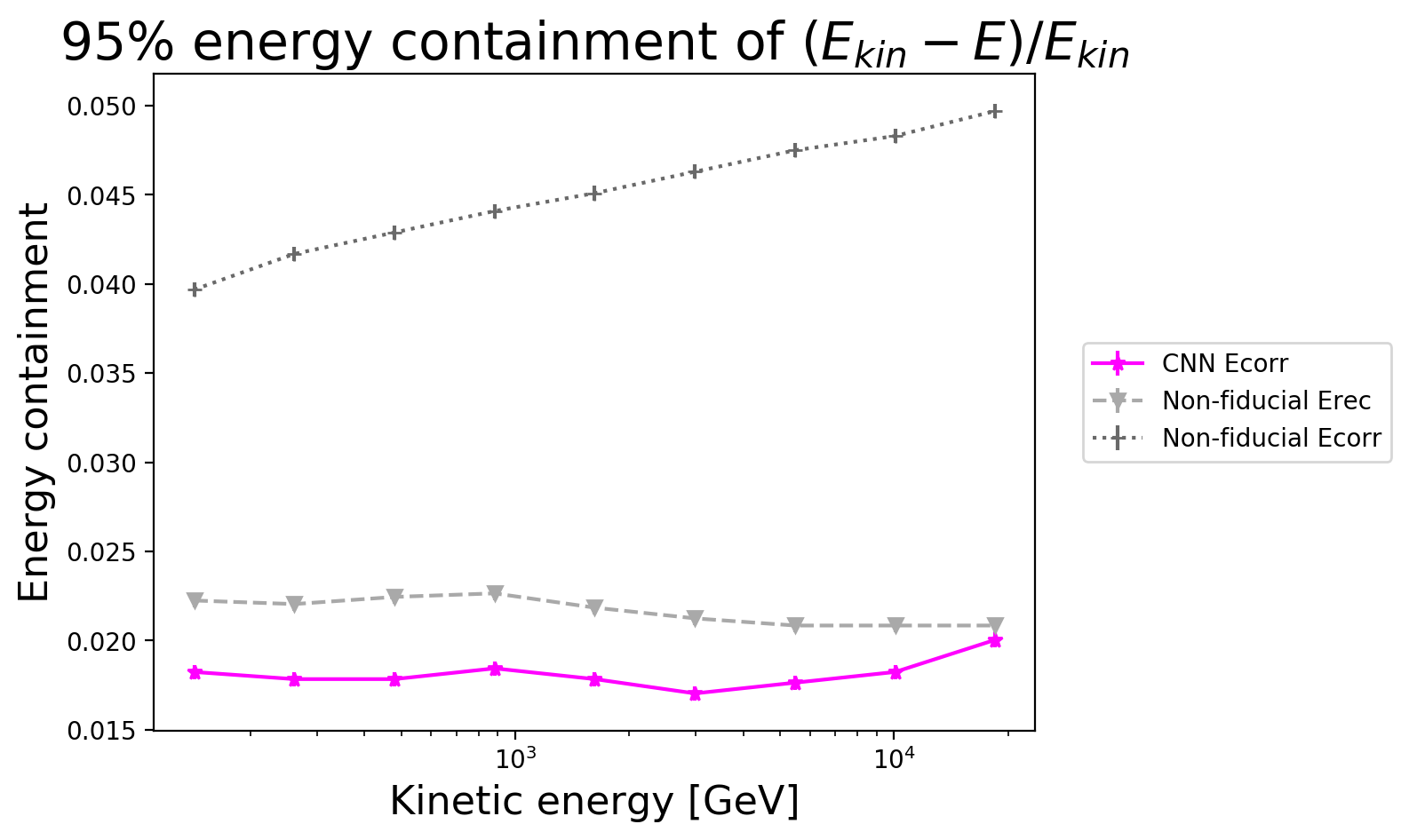}
\caption{The $68\%$ and $95\%$ containment of the relative energy reconstruction errors as function of the kinetic energy. \label{fig: radius ratio CNN}}
\end{figure}

From Figure~\ref{fig: ratio CNN}, we can see that the relative energy reconstruction error of the CNN model ($E_\mathrm{CNN}$) has a distribution centred close to 0, similar to $E_\mathrm{corr}$, indicating an improvement over $E_\mathrm{rec}$. Additionally, there is a decrease  in the number of events where the energy is underestimated compared to $E_\mathrm{kin}$. However, a slight bias toward negative values is observed, leading to a mild overestimation of the initial energy ($E_\mathrm{CNN}$>$E_\mathrm{kin}$). To quantify the efficiency of the CNN method with respect to other methods, we analyse the $68\%/ 95\%$ containment values of the distributions in Figure~\ref{fig: ratio CNN}, as shown in Figures~\ref{fig: mean ratio CNN} and \ref{fig: radius ratio CNN} respectively. Figure~\ref{fig: mean ratio CNN} shows that the mean value for the CNN method stays stable at all energy, demonstrating apparent improvement over the $E_\mathrm{corr}$ method. The same can be seen for $68\%/ 95\%$ containments in Figure~\ref{fig: radius ratio CNN}. In particular, the $68\%$ containment for the CNN method is consistently lower (better) that for the standard methods and decreases (improves) with energy. The $95\%$ containment shows similar behaviour although with less improvement compared to the standard methods, especially at high energies. These results demonstrate a significant improvement compared to the classical methods.

As we have shown, the CNN method allows to recover the missing shower energy in the calorimeter and consequently to have a close estimation of the real electron energy. 

\begin{figure}[htbp]
\centering
\includegraphics[width=.4\textwidth]{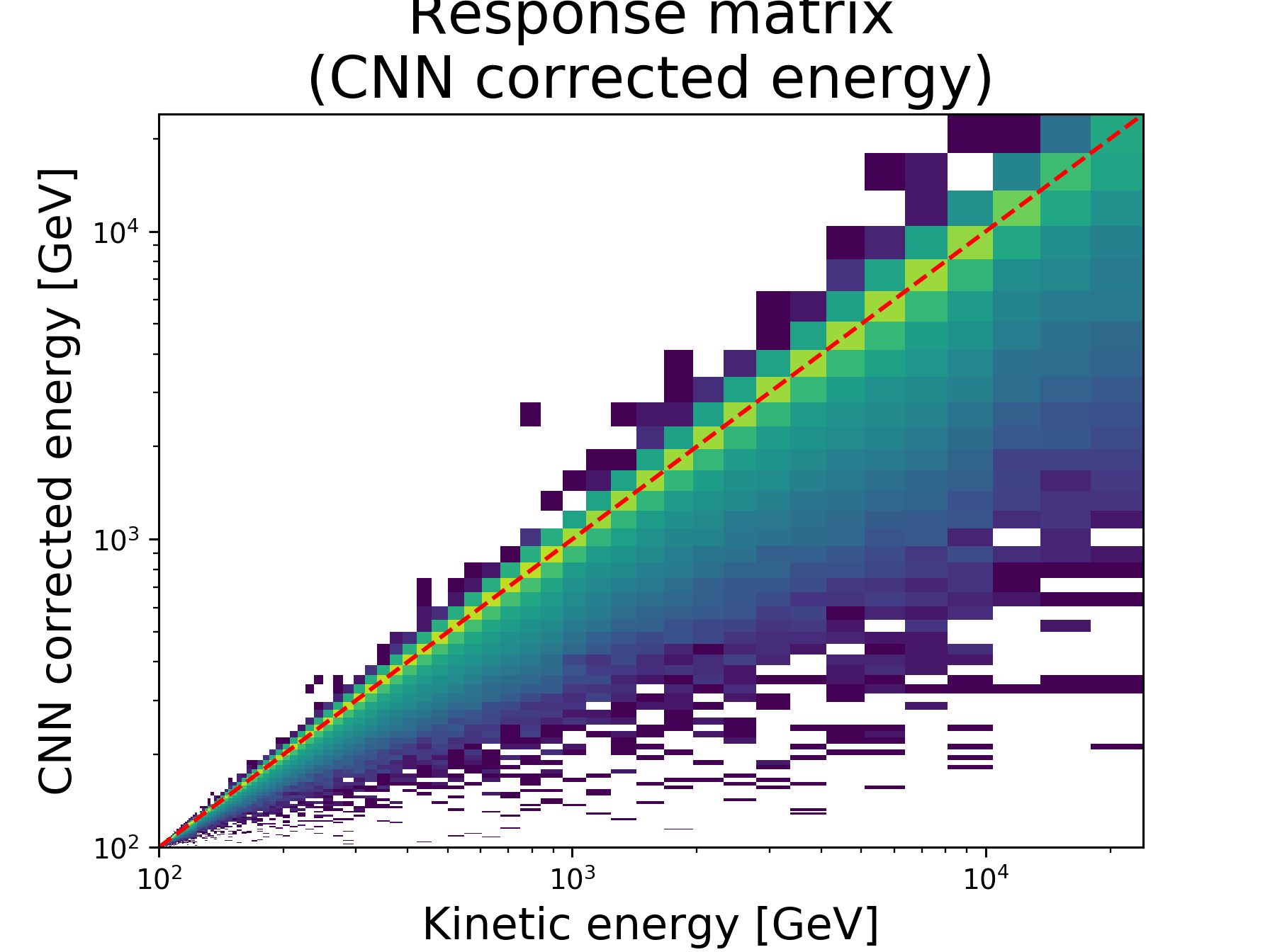}
\qquad
\includegraphics[width=.4\textwidth]{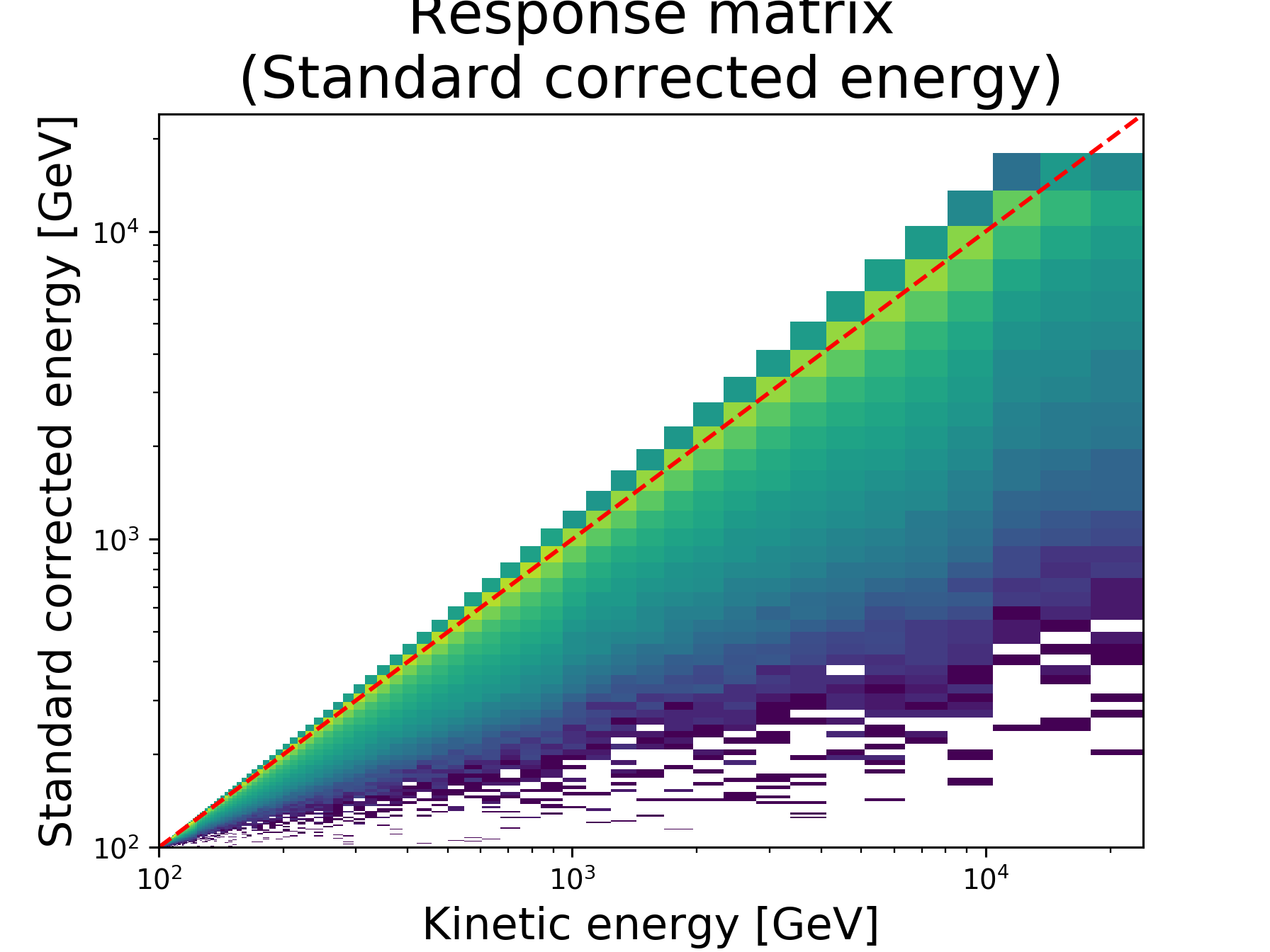}
\qquad
\includegraphics[width=.4\textwidth]{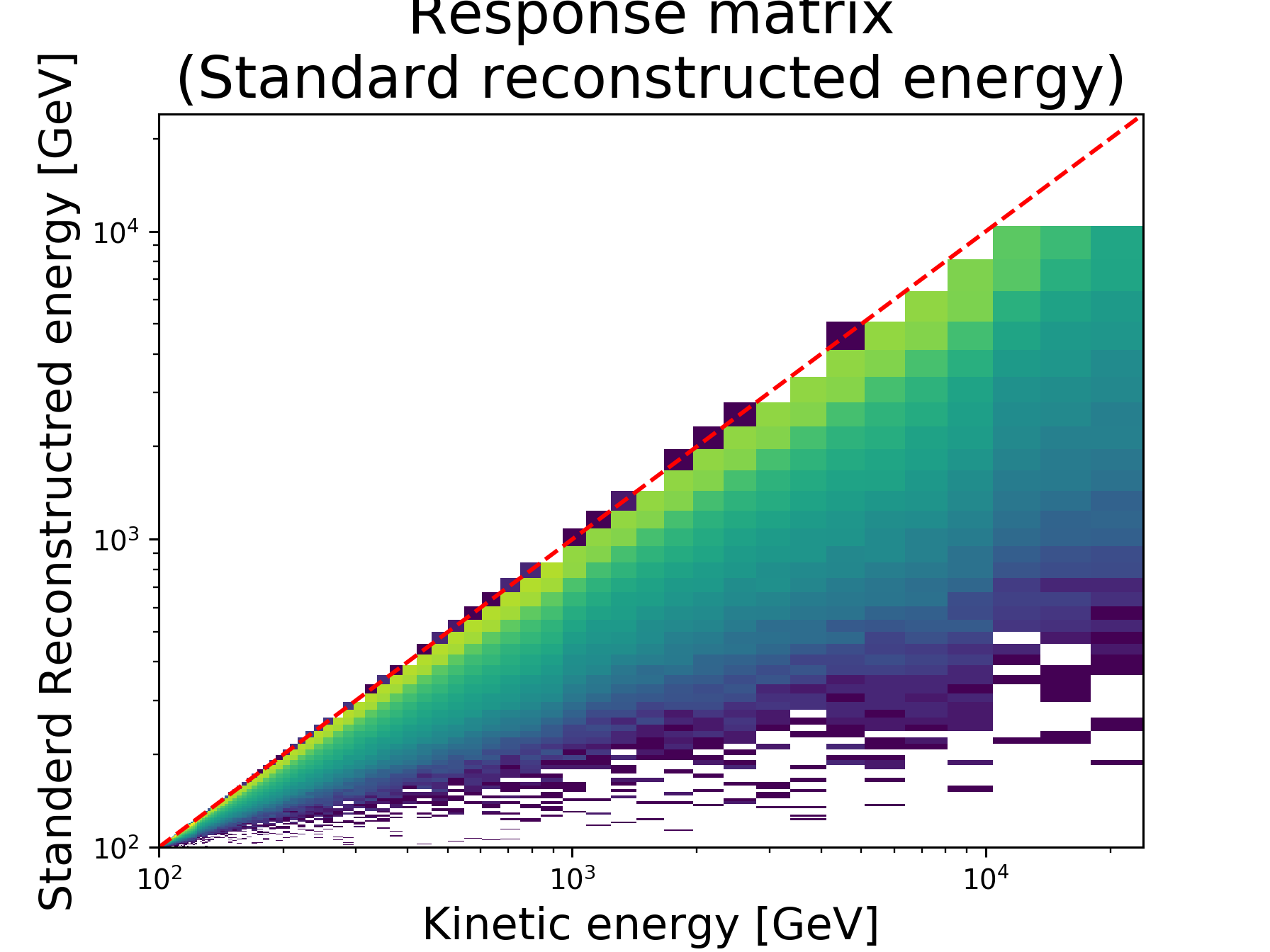}
\qquad

\caption{Response matrix of the three energy reconstruction methods: CNN-corrected, standard corrected and standard reconstruction (no correction). Red line indicates a diagonal of a matrix. \label{fig: response matrix}}
\end{figure}

%%This paragraph is new see with PAul
Unfolding \cite{DAGOSTINI1995487} is commonly used in the analysis of CR data to estimate the number of events as a function of the true energy by leveraging knowledge from the detector (typically using MC samples). This process involves constructing a response matrix that characterizes the deposited energy as a function of the kinetic energy (Figure~\ref{fig: response matrix}). We will demonstrate that the developed CNN method eliminates the need for unfolding in flux calculations.

To illustrate this, we will define a response matrix for each energy reconstruction method and use the Python library pyUnfold \cite{Bourbeau2018} to apply the unfolding method. We will then compare the obtained flux (after unfolding) to those derived without the unfolding step, showcasing the advantages of the CNN approach.

We can now compare the normalised number of events before and after unfolding.

\begin{figure}[htbp]
\centering
\includegraphics[scale=0.55]{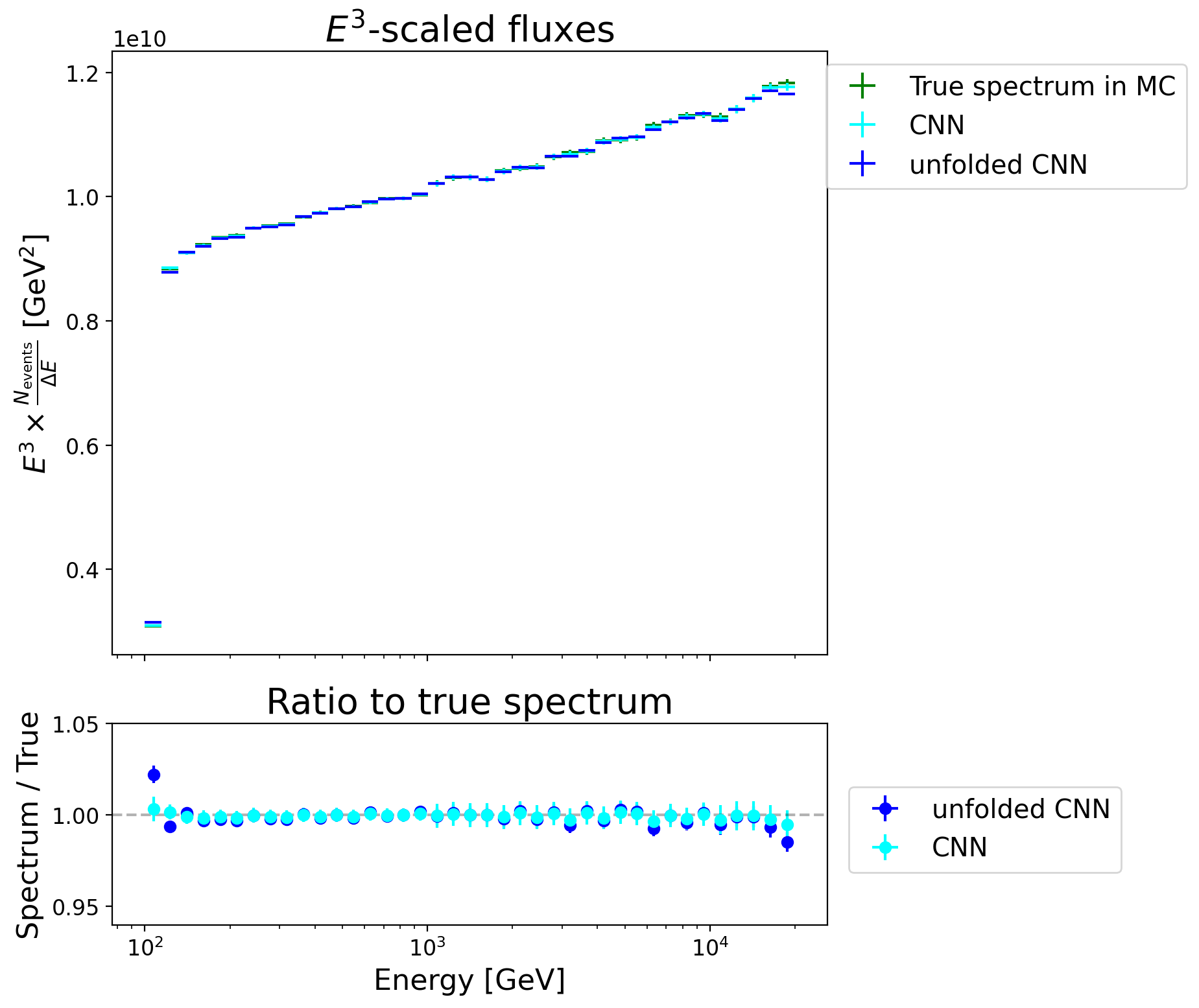}
\qquad

\caption{Electron kinetic energy spectrum obtained with the CNN method with and without the unfolding procedure (cyan and blue lines respectively) compared with the truth spectrum defined in MC. Bottom plots displays the ratio of the calculate spectrum, with and without unfolding, divided by the truth kinetic energy spectrum defined in MC.\label{fig: ratio unfold comparison}}
\end{figure}

From Figure~\ref{fig: ratio unfold comparison}, it can be seen that the normalized CNN energy is in good agreement with the kinetic energy, the same can be said for the unfolded CNN energy. The bottom plot from Figure~\ref{fig: ratio unfold comparison} shows the ratio of electron energy spectrum calculated using the CNN-corrected energy method, with and without the unfolding, divided by the truth spectrum defined in the MC simulation. We can see that the unfolding procedure does not improve the electron spectrum estimation compare to case without unfolding. In fact we can see an oscillation. On top of that, the unfolding procedure has an edge effect at low and high energies. 

We conclude that the CNN method can recover the real energy of the event accurately and with minimal bias. Because of this, unfolding is not needed to compute the electron flux. This is advantageous because it enables avoiding unfolding edge effects.

\section{Flight data comparison}
\label{section:Flight data comparison}
\begin{figure}%[htbp]
\centering
% Row 1

\begin{minipage}{0.45\textwidth}
    \centering
    \includegraphics[scale=0.15]{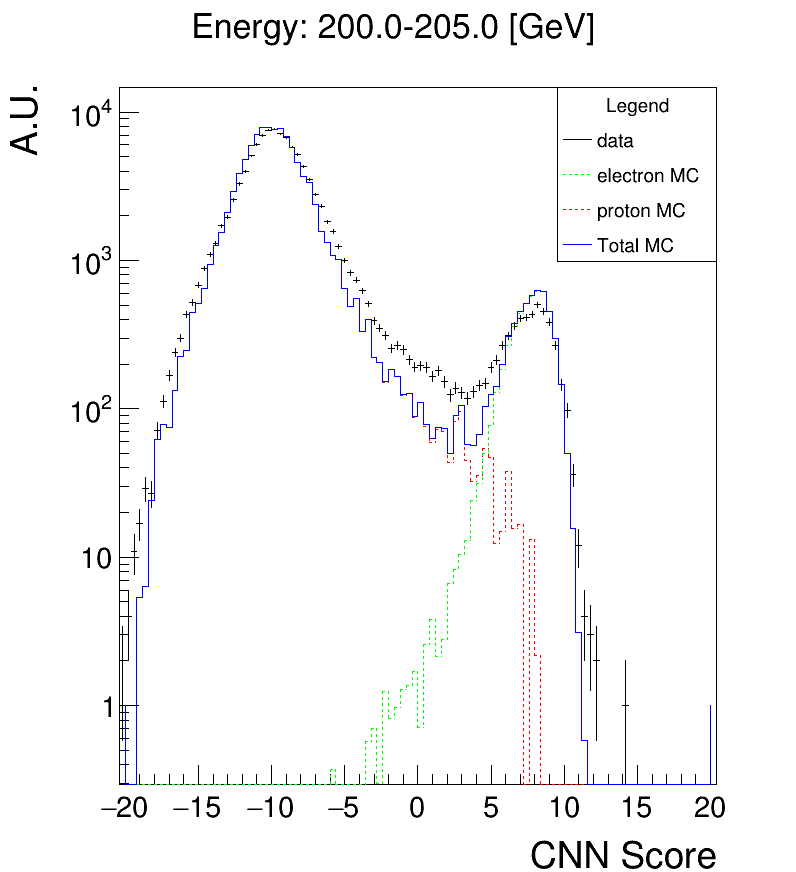}
\end{minipage}
\qquad
\begin{minipage}{0.45\textwidth}
    \centering
    \includegraphics[scale=0.15]{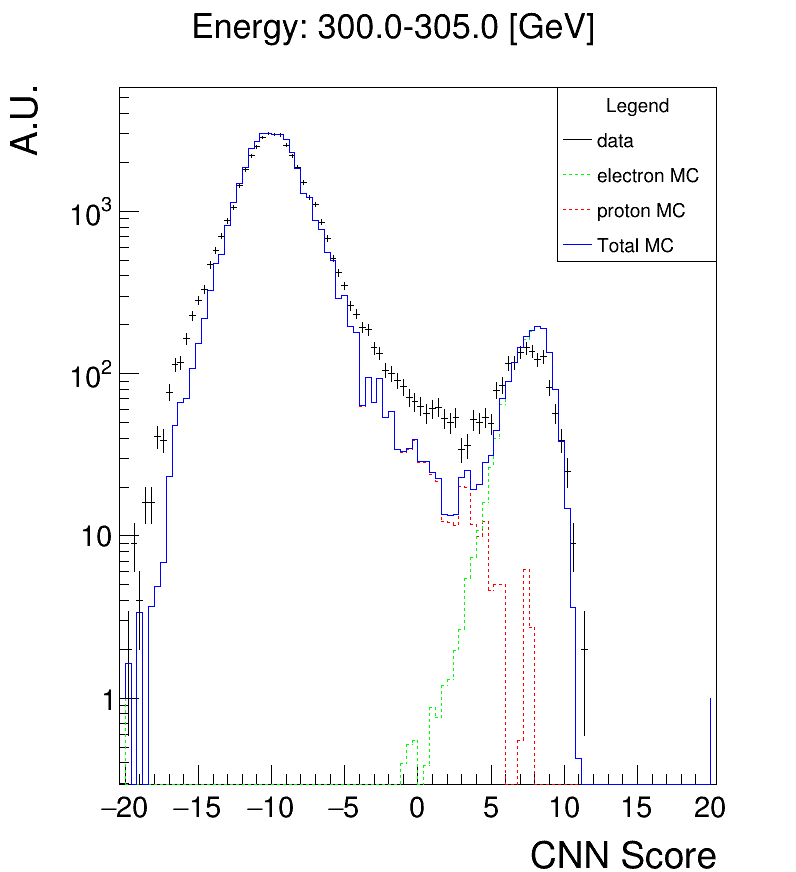}
\end{minipage}
\quad
%\vspace{0.5cm} % Vertical space between rows

% Row 2
\begin{minipage}{0.45\textwidth}
    \centering
    \includegraphics[scale=0.15]{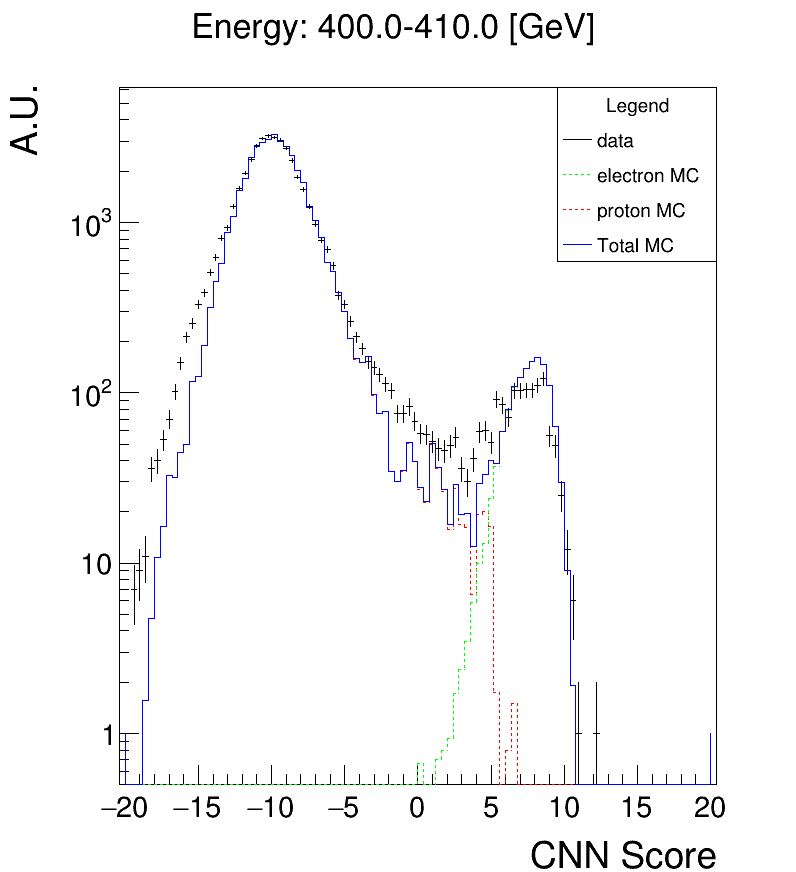}
\end{minipage}
\qquad
\begin{minipage}{0.45\textwidth}
    \centering
    \includegraphics[scale=0.15]{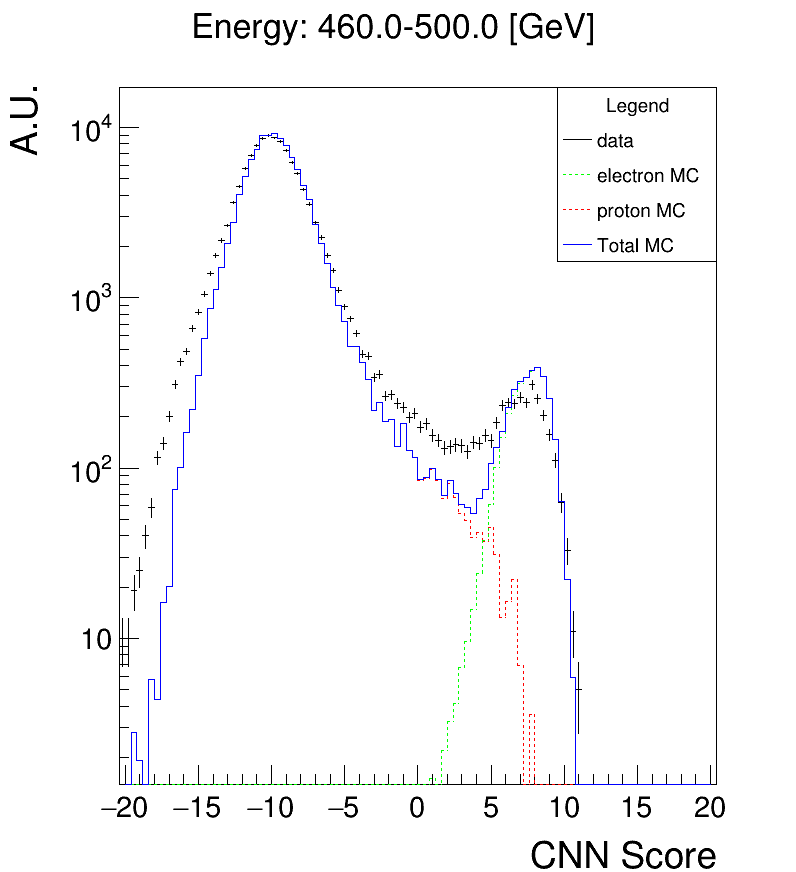}
\end{minipage}

%\vspace{0.5cm} % Vertical space between rows

% Row 3
\begin{minipage}{0.45\textwidth}
    \centering
    \includegraphics[scale=0.15]{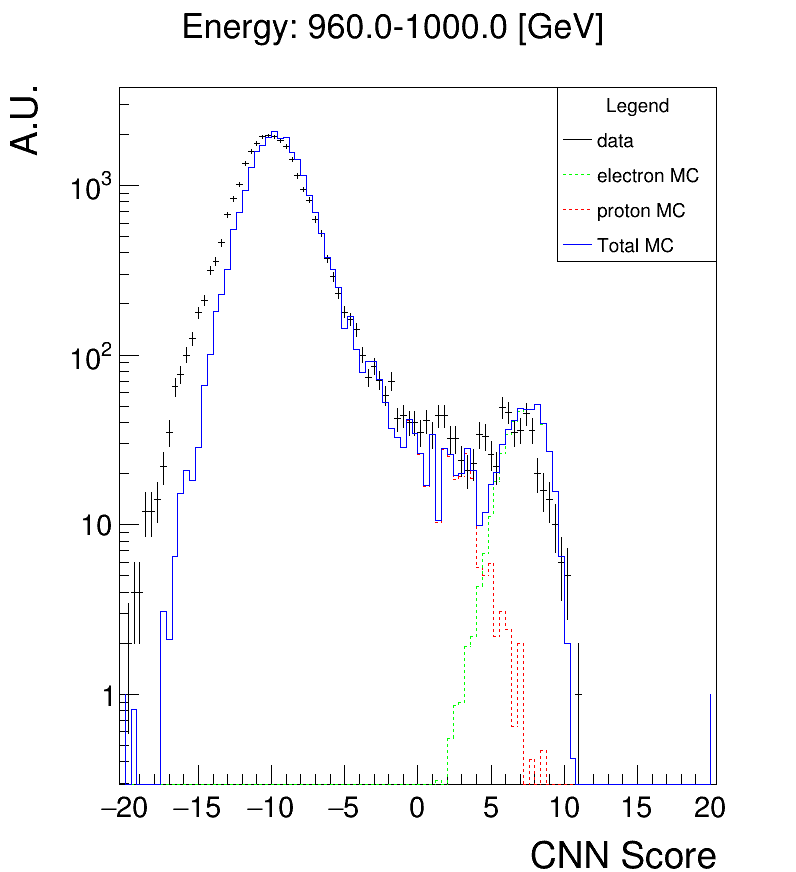}
\end{minipage}

\caption{Distribution of CNN classifier for data and MC samples for different energy bins in logarithmic scale.}
\label{fig: e-p cnn score log}

\end{figure}

\begin{figure}[htbp]
\centering
\includegraphics[width=.4\textwidth]{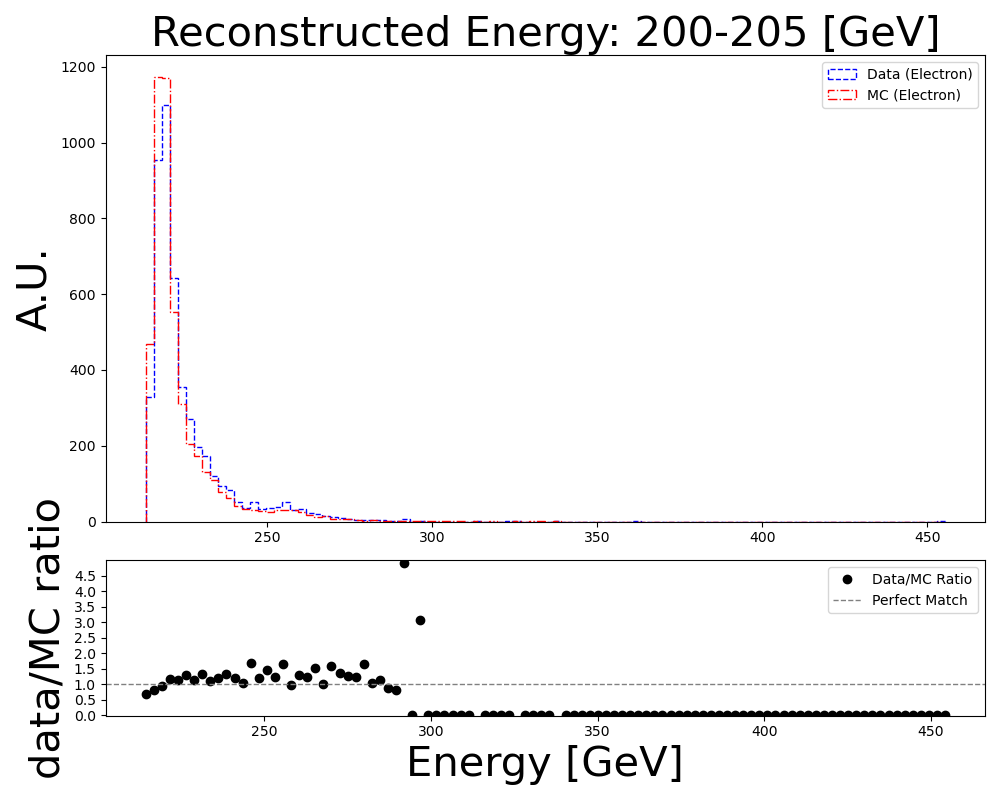}
\qquad
\includegraphics[width=.4\textwidth]{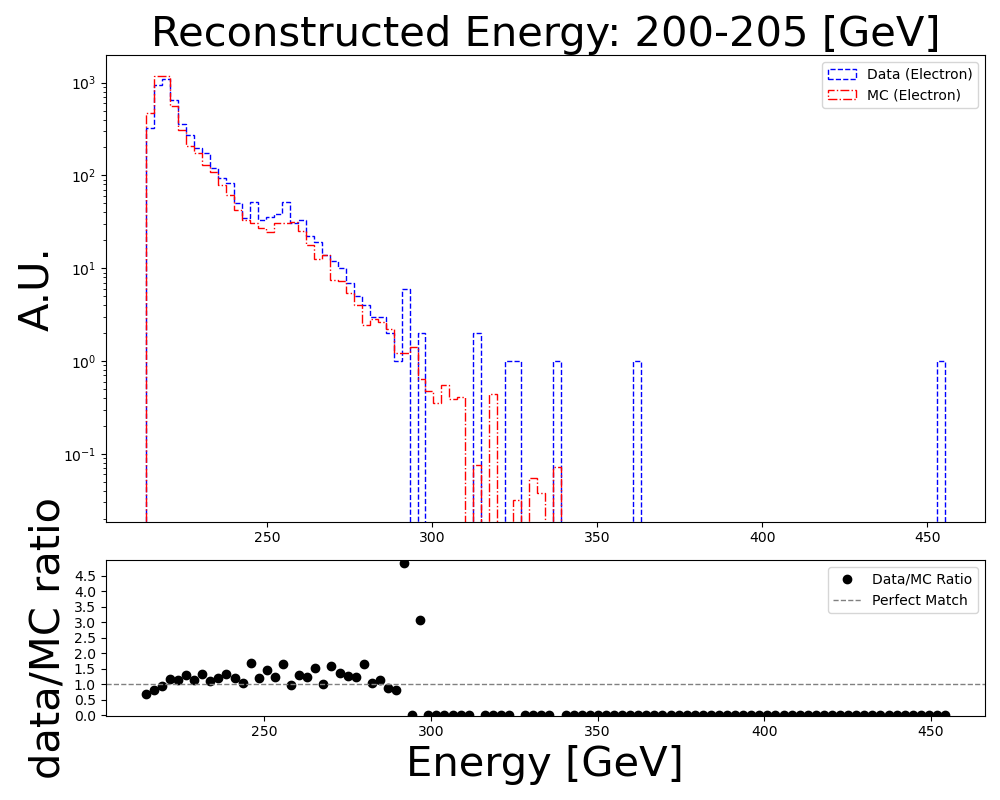}
\qquad
\includegraphics[width=.4\textwidth]{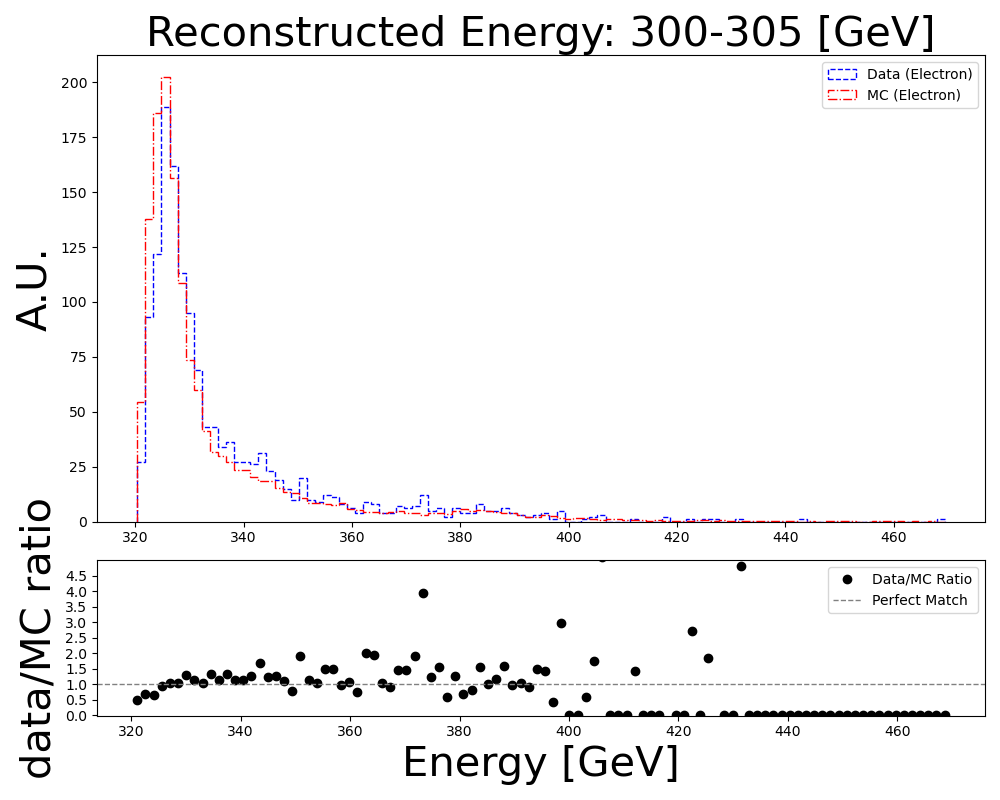}
\qquad
\includegraphics[width=.4\textwidth]{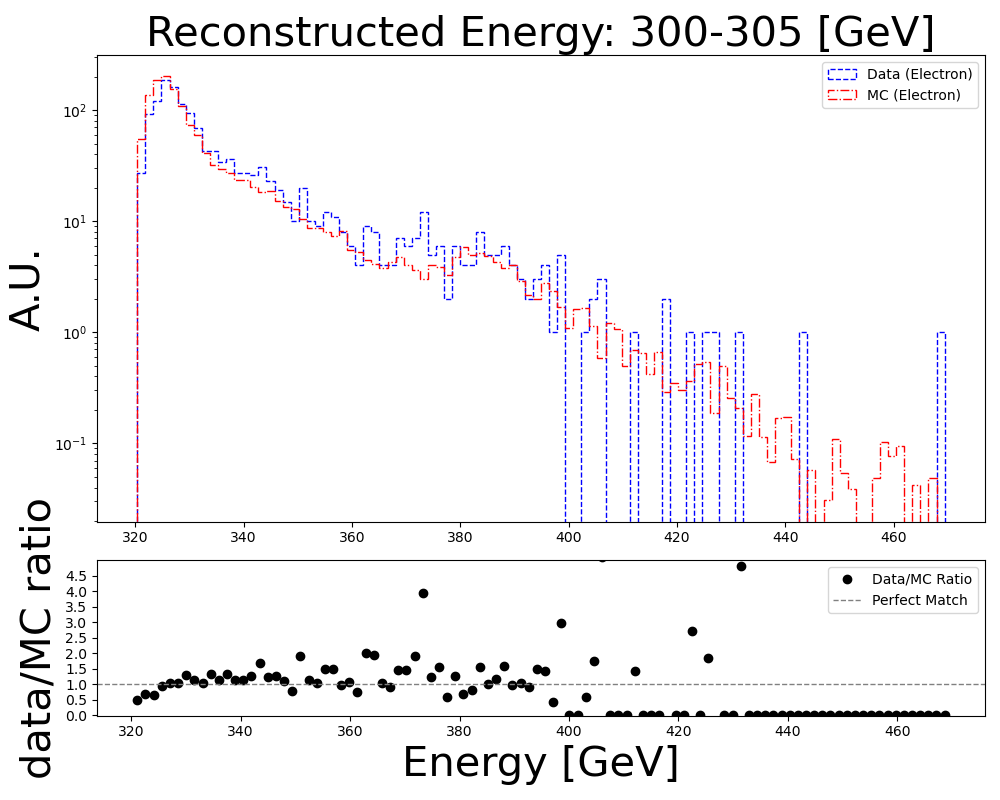}
\qquad
\includegraphics[width=.4\textwidth]{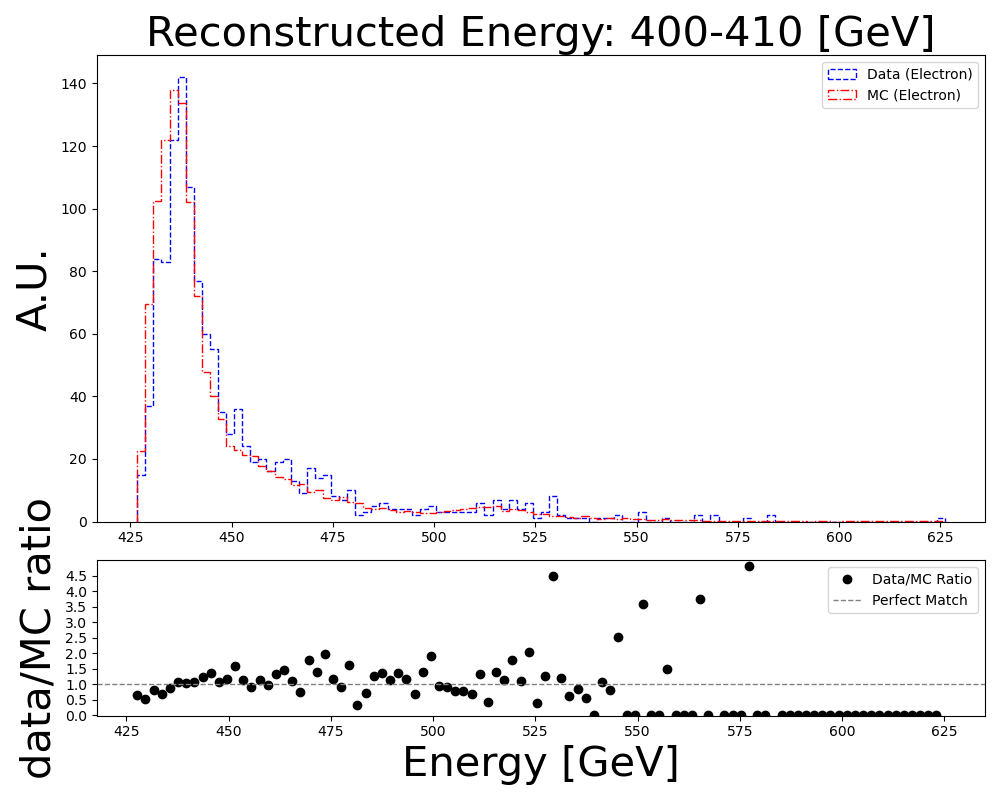}
\qquad
\includegraphics[width=.4\textwidth]{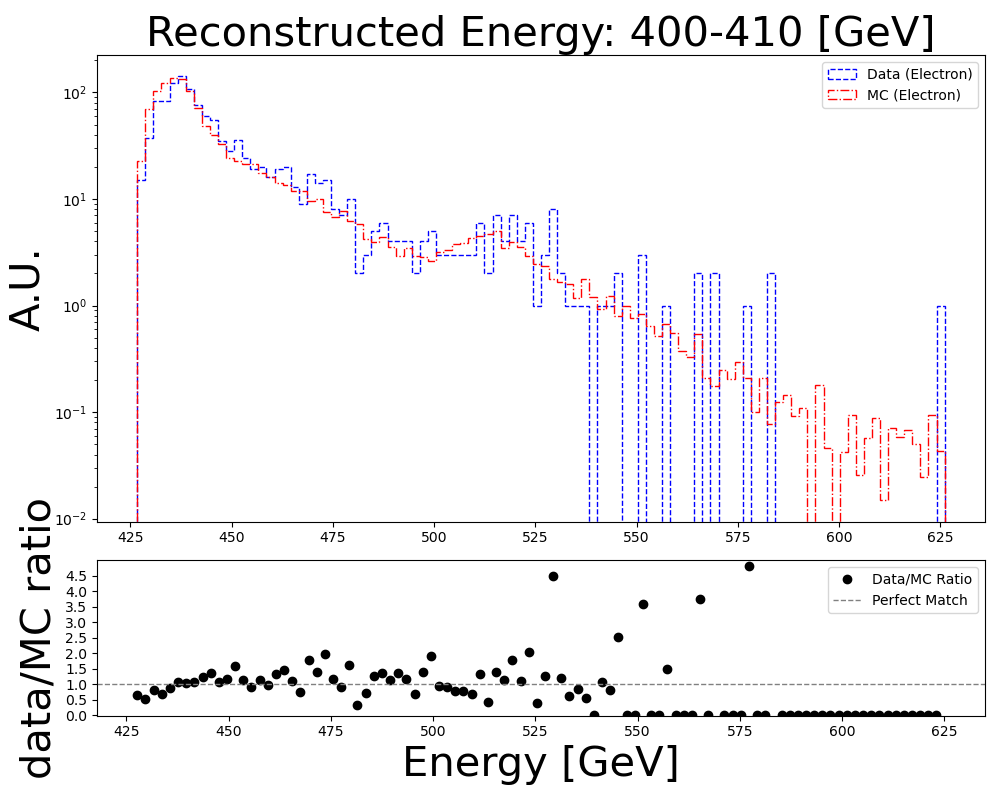}
\qquad
\includegraphics[width=.4\textwidth]{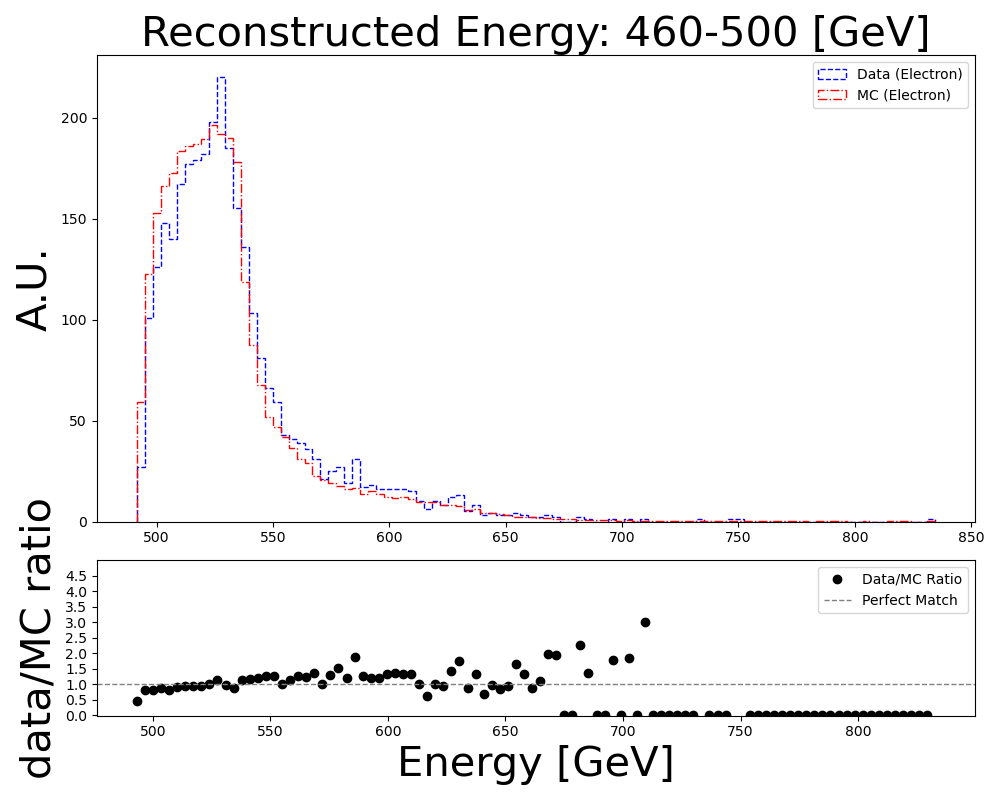}
\qquad
\includegraphics[width=.4\textwidth]{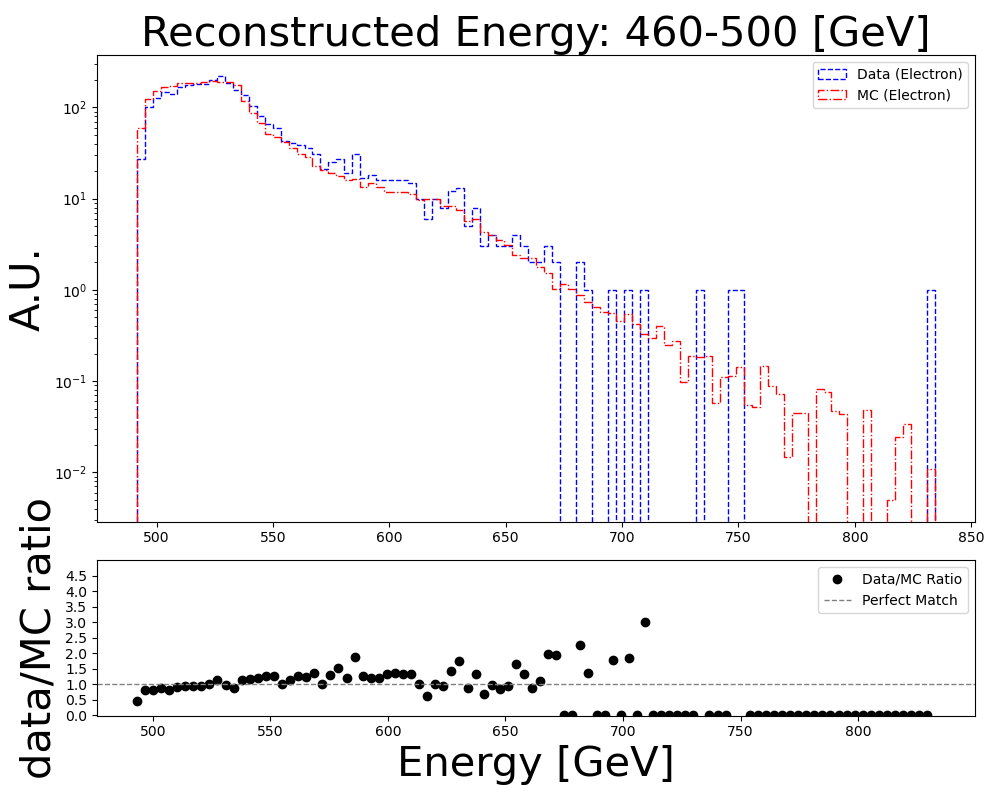}
\qquad
\includegraphics[width=.4\textwidth]{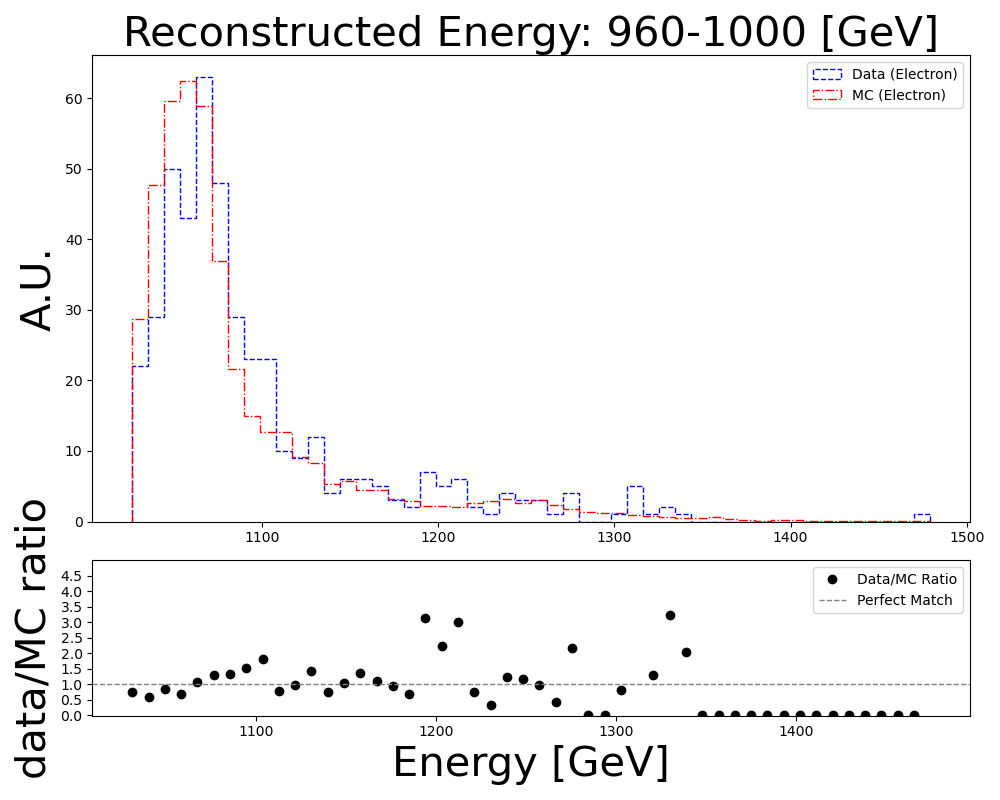}
\qquad
\includegraphics[width=.4\textwidth]{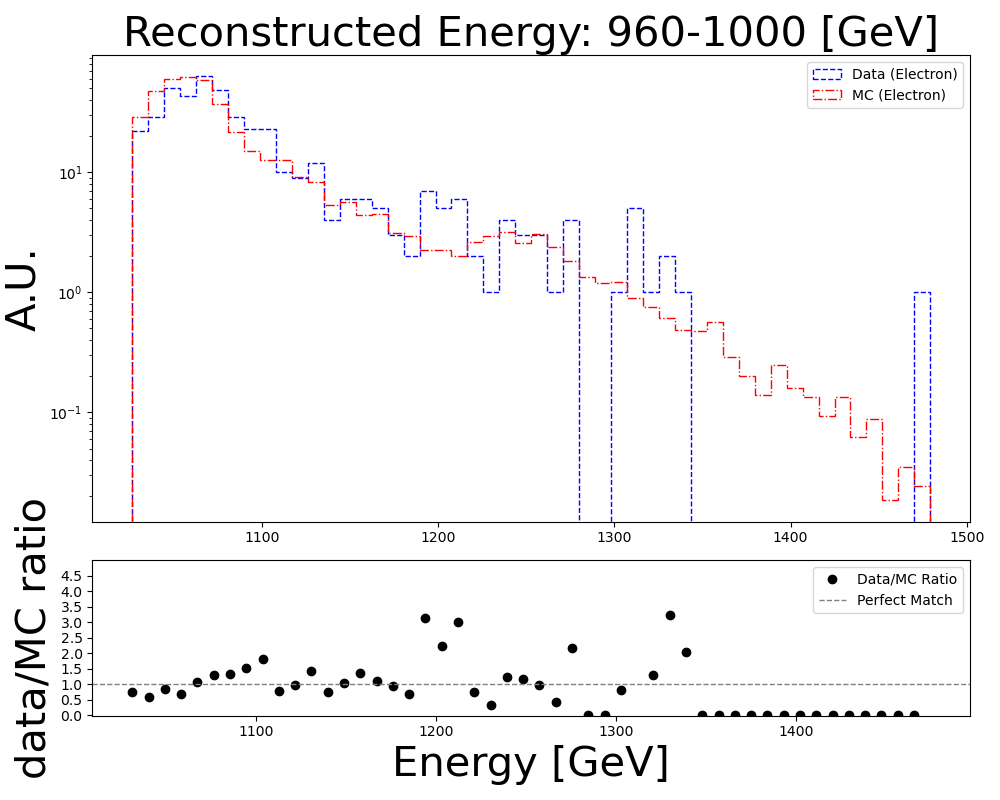}
\caption{Distribution of CNN energy correction method for data and MC samples. Lower plot are the ratio between real and MC data. Left plots are in linear scale, right plots are in log scale.\label{fig: 2A vs MC cnn}}
\end{figure}

In the previous section we looked at the efficiency of the CNN applied to MC  and compared it with the classical methods ($E_\mathrm{rec}$ and $E_\mathrm{corr}$). However, to be able to use such technique in future analyses, we need to confirm the previous observations for real data. 
To do so, we select a sample of events (data and MC) in a narrow energy range using the reconstructed energy since the kinetic energy is not accessible for real data. The electrons in real data are selected with another CNN, trained to separate electrons from protons. The training sample for this CNN was created using electrons and protons MC. Following \cite{Droz_ML}, we remove the sigmoid from the output layer, thus the score from the CNN results in a value between $[-\infty,+\infty]$, were negative values correspond to proton like events while positive values correspond to electron like events. Figure~\ref{fig: e-p cnn score log} show the output of the classifier for 5 energy samples divided in bins of $E_\mathrm{rec}$: 200-205 GeV, 300-305 GeV, 400-410 GeV, 460-500 GeV and 960-1000 GeV\footnote{The last two energy bins are larger due to low statistics.}, which represents the score values used to select a clean sample. A cut value of 4 is used to select  a clean sample. Table~\ref{tab:background fraction} shows the number of events that pass this cut. Finally, in Figure~\ref{fig: 2A vs MC cnn} we plot for each energy range the distribution of energy reconstructed with the CNN method for both real data and MC as shown.

\begin{figure}
\centering
\includegraphics[width=.5\textwidth]{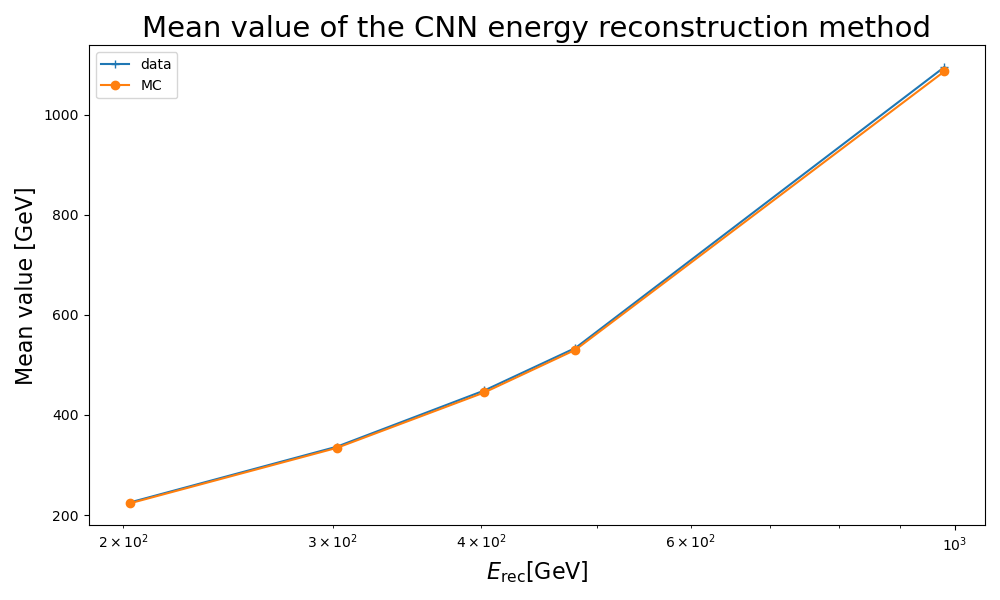}
\caption{Mean value of the CNN reconstructed energy for data and MC as a function of the standard energy reconstruction method \cite{CHANG20176}.\label{fig: mean 2A}}
\end{figure}

\begin{figure}
\centering
\includegraphics[width=.5\textwidth]{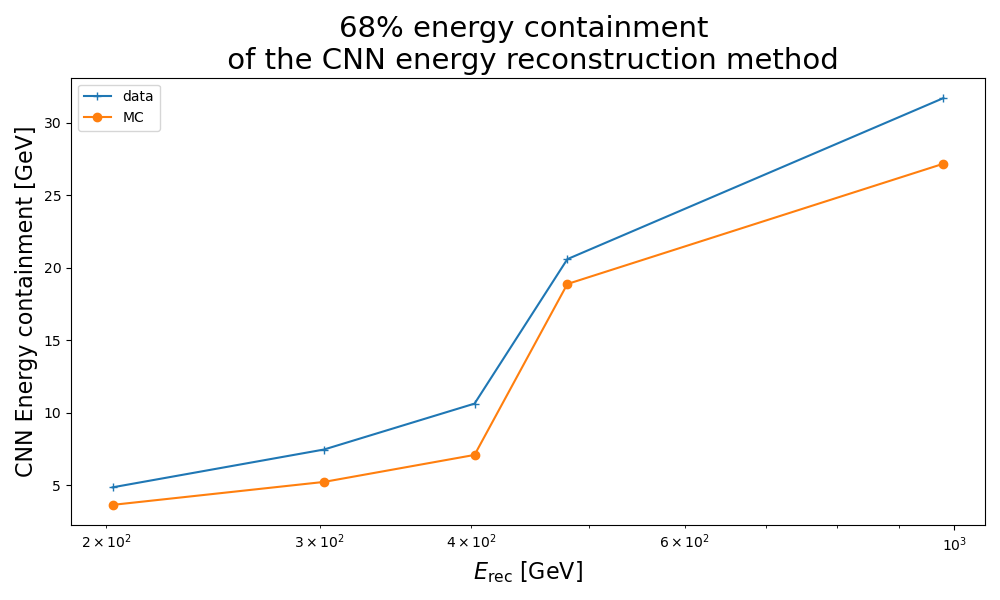}
\qquad
\includegraphics[width=.5\textwidth]{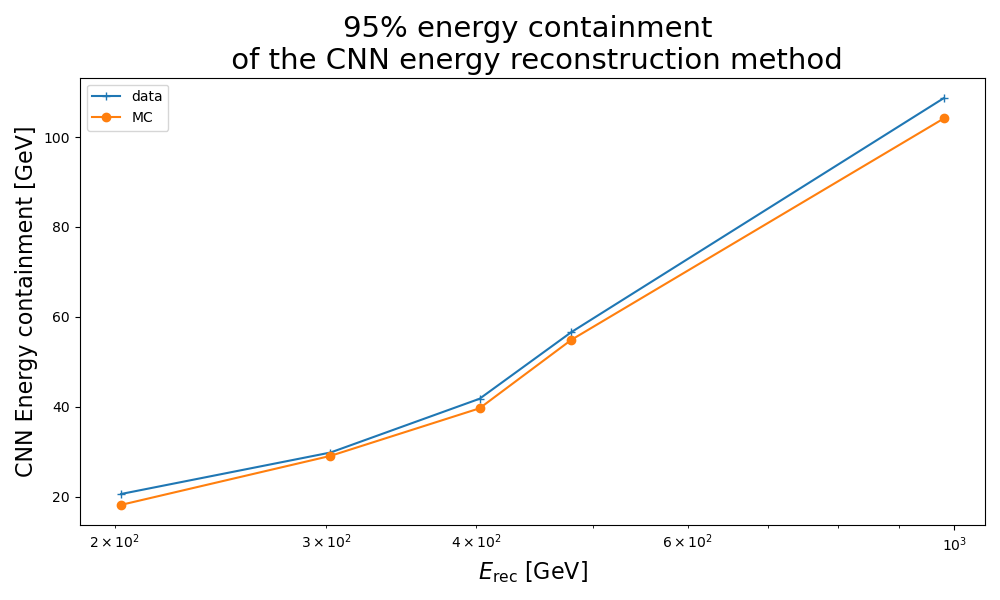}
\qquad
\caption{The 68\% and 95\% containment of the CNN energy reconstruction method as a function of standard energy reconstruction method \cite{CHANG20176}.\label{fig: 68.95 2A-MC}}
\end{figure}

In Figure~\ref{fig: 2A vs MC cnn} we can see an overall adequate agreement between real data and MC showing that the qualitative behaviour of the CNN is similar between data and MC, also shown by the ratio between data and MC. Adjusting the cut value of the CNN classifier to 3 worsens the agreement, while increasing it to 5 improves the match, suggesting that the background fraction plays a role in shaping the distributions shown in Figure~\ref{fig: 2A vs MC cnn}. Additionally, distinct features are visible in the tails of the distributions. These features arise because, for a given kinetic energy, a particle can deposit only a fraction of its energy, as illustrated by the response matrix in Figures~\ref{fig: response matrix}. The CNN's ability to more accurately estimate kinetic energy leads to the emergence of additional structures within the tail, corresponding to various kinetic energy values. To have a more quantitative way of showing the comparison between data and MC as shown in Figures~\ref{fig: mean 2A} and ~\ref{fig: 68.95 2A-MC}, we looked at the mean value and the 68/95 \% energy containment of the distributions shown in Figure~\ref{fig: 2A vs MC cnn}. The comparison between data and MC is performed using two metrics: the mean value and the 68/95\% energy containment. The results show a strong agreement in the mean values between data and MC, indicating consistency in the central tendencies. However, the energy containment demonstrates tighter bounds in the MC compared to the real data. Despite this difference, both data and MC exhibit similar overall behaviour, as illustrated in Figures~\ref{fig: mean 2A} and ~\ref{fig: 68.95 2A-MC}, which validate the alignment of the models in capturing key trends.

\begin{table}[htbp]
    \centering
    \caption{Number of MC and data events passing the CNN classifier, for a cut value of 4. Last column represents the fraction of background that passes the cut.}
    
    \begin{tabular}{|l|l|l|l|l|}
    \hline
        Energy range [GeV] & proton MC & electron MC & data & background fraction\\  \hline
         200-205 & 279 & 4763 & 4793 & 5.8$\%$ \\ \hline
         300-305 & 63 & 1508 & 1429 & 4.4$\%$\\ \hline
         400-410 & 63 & 1246  & 1215 & 4.5 $\%$ \\ \hline
         460-500 & 254 & 3045 & 2988 & 8.5 $\%$\\ \hline
         960-1000 & 49 & 401 & 434 & 11.3 $\%$ \\ \hline
    \end{tabular}
    
    \label{tab:background fraction}
\end{table}


\section{Conclusion}
\label{section: Conclusion}
Studies suggest the potential for the direct observation of cosmic ray electron (CRE) and positron sources at TeV scale \cite{CREs_origin,2004ApJ...601..340K}. Additionally, various dark matter models predict electron and positron production through annihilation or decay \cite{DM_model}, which may manifest as an excess in the high energy CRE spectrum. However, due to their light mass, CREs experience significant energy losses during propagation. Combined with an already lower flux compared to the other particles species, CREs are relatively rare at high energies. 
To increase statistics and extend DAMPE's current results \cite{DAMPE_elec_2017}, the inclusion of non-fiducial electrons and positron is necessary. However, incorporating these events introduces challenges due to their complex shower topology in the calorimeter. One such challenge, addressed in this paper, is their accurate energy reconstruction. As demonstrated, classical methods are not well-suited for non-fiducial events (see Figures~\ref{fig: ratio Erec and Ecorr},\ref{fig: mean ratio},\ref{fig: radius ratio}). Therefore, the application of a Convolutional Neural Network (CNN) for regression offers a promising solution.\\
The training process for the CNN involved splitting the Monte Carlo sample into training (70$\%$) and validation (30$\%$) sets, with input images from the BGO calorimeter. The regression was performed on the relative error between the reconstructed energy and the true kinetic energy. Figure~\ref{fig: CNN loss} shows the evolution of the CNN’s performance, measured via the mean squared error (Equation~\ref{Eq: mean squared error}). We evaluated the CNN’s effectiveness by comparing its performance to classical methods, focusing on the relative error distributions (Figure~\ref{fig: ratio CNN}). Additionally, we quantified the CNN’s performance. The mean values (Figure~\ref{fig: mean ratio CNN}) have shown stable results at all energies (close to a mean of 0). The 68$\%$ and 95$\%$ energy resolution plots (Figure~\ref{fig: radius ratio CNN}) show an improvement compared to the classical methods. Then, we considered unfolding, a step that is usually performed during a flux analysis, but our results demonstrated that the CNN approach eliminates the need for unfolding, avoiding the introduction of edge effects as shown in Figure~\ref{fig: ratio unfold comparison}. 

Although MC demonstrates the improvement of the CNN method over the classical approaches, it is essential to validate these results with data. To achieve this, we used an ML classifier, more precisely a CNN, to select a clean sample, from data, of electrons and divided them into five samples based on energy ranges, [200-205, 300-305, 400-410, 460-500, 960-1000] GeV. We then compared it to a MC sample selected using the same criteria. Figure~\ref{fig: 2A vs MC cnn} shows good qualitative agreement between the two, as estimated by the ratio of both distributions. Similar conclusions can be reached by looking at the mean of distributions  and the 68/95\% energy containment for real data and MC. In conclusion, the CNN method offers a significant improvement in accuracy and precision of estimating/recovering the initial energy.
\newpage
%% [A] Recommended: using JHEP.bst file
\bibliographystyle{JHEP}
\bibliography{biblio.bib}

\providecommand{\href}[2]{#2}\begingroup\raggedright\begin{thebibliography}{10}

\bibitem{CHANG20176}
J.~Chang, G.~Ambrosi, Q.~An, R.~Asfandiyarov, P.~Azzarello, P.~Bernardini et~al., \emph{The dark matter particle explorer mission}, \href{https://doi.org/https://doi.org/10.1016/j.astropartphys.2017.08.005}{\emph{Astroparticle Physics} {\bfseries 95} (2017) 6}.

\bibitem{YU20171}
Y.~Yu, Z.~Sun, H.~Su, Y.~Yang, J.~Liu, J.~Kong et~al., \emph{The plastic scintillator detector for dampe}, \href{https://doi.org/https://doi.org/10.1016/j.astropartphys.2017.06.004}{\emph{Astroparticle Physics} {\bfseries 94} (2017) 1}.

\bibitem{AZZARELLO2016378}
P.~Azzarello, G.~Ambrosi, R.~Asfandiyarov, P.~Bernardini, B.~Bertucci, A.~Bolognini et~al., \emph{The dampe silicon–tungsten tracker}, \href{https://doi.org/https://doi.org/10.1016/j.nima.2016.02.077}{\emph{Nuclear Instruments and Methods in Physics Research Section A: Accelerators, Spectrometers, Detectors and Associated Equipment} {\bfseries 831} (2016) 378}.

\bibitem{WEI2019177}
Y.~Wei, Y.~Zhang, Z.~Zhang, L.~Wu, S.~Wen, H.~Dai et~al., \emph{Performance of the dampe bgo calorimeter on the ion beam test}, \href{https://doi.org/https://doi.org/10.1016/j.nima.2018.12.036}{\emph{Nuclear Instruments and Methods in Physics Research Section A: Accelerators, Spectrometers, Detectors and Associated Equipment} {\bfseries 922} (2019) 177}.

\bibitem{Huang_2020}
Y.-Y.~Huang, T.~Ma, C.~Yue, Y.~Zhang, M.-S.~Cai, J.~Chang et~al., \emph{Calibration and performance of the neutron detector onboard of the dampe mission}, \href{https://doi.org/10.1088/1674-4527/20/9/153}{\emph{Research in Astronomy and Astrophysics} {\bfseries 20} (2020) 153}.

\bibitem{PhysRev.137.B1306}
F.C.~Jones, \emph{Inverse compton scattering of cosmic-ray electrons}, \href{https://doi.org/10.1103/PhysRev.137.B1306}{\emph{Phys. Rev.} {\bfseries 137} (1965) B1306}.

\bibitem{CREs_origin}
C.~Evoli, E.~Amato, P.~Blasi and R.~Aloisio, \emph{Galactic factories of cosmic-ray electrons and positrons}, \href{https://doi.org/10.1103/PhysRevD.103.083010}{\emph{Phys. Rev. D} {\bfseries 103} (2021) 083010}.

\bibitem{2004ApJ...601..340K}
T.~{Kobayashi}, Y.~{Komori}, K.~{Yoshida} and J.~{Nishimura}, \emph{{The Most Likely Sources of High-Energy Cosmic-Ray Electrons in Supernova Remnants}}, \href{https://doi.org/10.1086/380431}{\emph{apj} {\bfseries 601} (2004) 340}.

\bibitem{DM_model}
M.S.~Turner and F.~Wilczek, \emph{Positron line radiation as a signature of particle dark matter in the halo}, \href{https://doi.org/10.1103/PhysRevD.42.1001}{\emph{Physical Review D} {\bfseries 42} (1990) 1001}.

\bibitem{Conrad2017}
J.~Conrad and O.~Reimer, \emph{Indirect dark matter searches in gamma and cosmic rays}, \href{https://doi.org/10.1038/nphys4049}{\emph{Nature Physics} {\bfseries 13} (2017) 224}.

\bibitem{PhysRevLett.122.041102}
{\scshape AMS Collaboration} collaboration, \emph{Towards understanding the origin of cosmic-ray positrons}, \href{https://doi.org/10.1103/PhysRevLett.122.041102}{\emph{Phys. Rev. Lett.} {\bfseries 122} (2019) 041102}.

\bibitem{PhysRevD.79.041301}
M.~Pohl, \emph{Cosmic-ray electron signatures of dark matter}, \href{https://doi.org/10.1103/PhysRevD.79.041301}{\emph{Phys. Rev. D} {\bfseries 79} (2009) 041301}.

\bibitem{DAMPE_elec_2017}
{DAMPE Collaboration}, \emph{Direct detection of a break in the teraelectronvolt cosmic-ray spectrum of electrons and positrons}, \href{https://doi.org/10.1038/nature24475}{\emph{Nature} {\bfseries 552} (2017) 63}.

\bibitem{BGO_correction}
C.~Yue, J.~Zang, T.~Dong, X.~Li, Z.~Zhang, S.~Zimmer et~al., \emph{A parameterized energy correction method for electromagnetic showers in bgo-ecal of dampe}, \href{https://doi.org/https://doi.org/10.1016/j.nima.2017.03.013}{\emph{Nuclear Instruments and Methods in Physics Research Section A: Accelerators, Spectrometers, Detectors and Associated Equipment} {\bfseries 856} (2017) 11}.

\bibitem{Droz_ML}
D.~Droz, A.~Tykhonov, X.~Wu, F.~Alemanno, G.~Ambrosi, E.~Catanzani et~al., \emph{A neural network classifier for electron identification on the dampe experiment}, \href{https://doi.org/10.1088/1748-0221/16/07/P07036}{\emph{Journal of Instrumentation} {\bfseries 16} (2021) }.

\bibitem{ML_allPhysics}
G.~Carleo, I.~Cirac, K.~Cranmer, L.~Daudet, M.~Schuld, N.~Tishby et~al., \emph{Machine learning and the physical sciences}, \href{https://doi.org/10.1103/RevModPhys.91.045002}{\emph{Reviews of Modern Physics} {\bfseries 91} (2019) 045002}.

\bibitem{PCA}
Z.~Xu, X.~Li, M.~Cui, C.~Yue, W.~Jiang, W.~Li et~al., \emph{An unsupervised machine learning method for electron–proton discrimination of the dampe experiment}, \href{https://doi.org/10.3390/universe8110570}{\emph{Universe} {\bfseries 8} (2022) }.

\bibitem{Andrii_ML}
A.~Tykhonov, A.~Kotenko, P.~Coppin, M.~Deliyergiyev, D.~Droz, J.M.~Frieden et~al., \emph{A deep learning method for the trajectory reconstruction of cosmic rays with the dampe mission}, \href{https://doi.org/https://doi.org/10.1016/j.astropartphys.2022.102795}{\emph{Astroparticle Physics} {\bfseries 146} (2023) 102795}.

\bibitem{Misha_ML}
M.~Stolpovskiy, X.~Wu, A.~Tykhonov, M.~Deliyergiyev, C.~Perrina, M.M.~Salinas et~al., \emph{Machine learning-based method of calorimeter saturation correction for helium flux analysis with dampe experiment}, \href{https://doi.org/10.1088/1748-0221/17/06/P06031}{\emph{Journal of Instrumentation} (2022) }.

\bibitem{AGOSTINELLI2003250}
S.~Agostinelli, J.~Allison, K.~Amako, J.~Apostolakis, H.~Araujo, P.~Arce et~al., \emph{Geant4—a simulation toolkit}, \href{https://doi.org/https://doi.org/10.1016/S0168-9002(03)01368-8}{\emph{Nuclear Instruments and Methods in Physics Research Section A: Accelerators, Spectrometers, Detectors and Associated Equipment} {\bfseries 506} (2003) 250}.

\bibitem{tensorflow2015-whitepaper}
M.~Abadi, A.~Agarwal, P.~Barham, E.~Brevdo, Z.~Chen, C.~Citro et~al., \emph{{TensorFlow}: Large-scale machine learning on heterogeneous systems},  2015.

\bibitem{Simonyan15}
K.~Simonyan and A.~Zisserman, \emph{Very deep convolutional networks for large-scale image recognition},  in \emph{International Conference on Learning Representations}, 2015.

\bibitem{NIPS2012_c399862d}
A.~Krizhevsky, I.~Sutskever and G.E.~Hinton, \emph{Imagenet classification with deep convolutional neural networks},  in \emph{Advances in Neural Information Processing Systems}, F.~Pereira, C.~Burges, L.~Bottou and K.~Weinberger, eds., vol.~25, Curran Associates, Inc., 2012.

\bibitem{ParticleDataGroup:2024cfk}
{\scshape Particle Data Group} collaboration, \emph{{Review of particle physics}}, \href{https://doi.org/10.1103/PhysRevD.110.030001}{\emph{Phys. Rev. D} {\bfseries 110} (2024) 030001}.

\bibitem{DAGOSTINI1995487}
G.~D'Agostini, \emph{A multidimensional unfolding method based on bayes' theorem}, \href{https://doi.org/https://doi.org/10.1016/0168-9002(95)00274-X}{\emph{Nuclear Instruments and Methods in Physics Research Section A: Accelerators, Spectrometers, Detectors and Associated Equipment} {\bfseries 362} (1995) 487}.

\bibitem{Bourbeau2018}
J.~Bourbeau and Z.~Hampel-Arias, \emph{Pyunfold: A python package for iterative unfolding}, \href{https://doi.org/10.21105/joss.00741}{\emph{The Journal of Open Source Software} {\bfseries 3} (2018) 741}.

\end{thebibliography}\endgroup

%% or
%% [B] Manual formatting (see below)
%% (i) We suggest to always provide author, title and journal data or doi:
%% in short all the informations that clearly identify a document.
%% (ii) please avoid comments such as "For a review'', "For some examples",
%% "and references therein" or move them in the text. In general, please leave only references in the bibliography and move all
%% accessory text in footnotes.
%% (iii) Also, please have only one work for each \bibitem.

\end{document}